\documentclass[preprint,eqsecnum,preprintnumbers,nofootinbib,byrevtex,prd,aps,showpacs,showkeys,groupedaddress,floatfix]{revtex4}
\usepackage{bm}
\usepackage{graphics}
\usepackage{graphicx}
\usepackage{epsfig}
\usepackage{amssymb}
\usepackage{amsmath}
\begin{document}
\title{INFRARED RENORMALONS AND SINGLE MESON PRODUCTION IN PROTON-PROTON COLLISIONS}
\author{A.~I.~Ahmadov$^{1}$~\footnote{ahmadovazar@yahoo.com}}
\author{Coskun~Aydin$^{3}$~\footnote{coskun@ktu.edu.tr}}
\author{Sh. M. Nagiyev$^{2}$}
\author{Yilmaz A.~Hakan$^{3}$~\footnote{hakany@ktu.edu.tr}}
\author{E.~A.~Dadashov$^{2}$}
\affiliation {$^{1}$ Institute for Physical Problems, Baku State
University, Z. Khalilov Street 23, AZ-1148, Baku, Azerbaijan \\
$^{2}$Institute of Physics of Azerbaijan National Academy of
Sciences, H. Javid Avenue, 33, AZ-1143, Baku, Azerbaijan \\
$^{3}$ Department of Physics, Karadeniz Technical University,
Trabzon, Turkey}
\begin{abstract} In this article, we investigate the contribution of
the higher-twist Feynman diagrams to the large-$p_T$  inclusive pion
production cross section in proton-proton collisions and present the
general formulae for the higher-twist differential cross sections in
the case of the running coupling and frozen coupling approaches. The
structure of infrared renormalon singularities of the higher twist
subprocess cross section and the resummed expression (the Borel sum)
for it are found. We compared the resummed higher-twist cross
sections with the ones obtained in the framework of the frozen
coupling approach and leading-twist cross section. We obtain, that
ratio
$R=(\Sigma_{\pi^{+}}^{HT})^{res}$/$(\Sigma_{\pi^{+}}^{HT})^{0}$, for
all values of the transverse momentum $p_{T}$ of the pion
identically equivalent to ratio
$r$=($\Delta_{\pi}^{HT})^{res}$/$(\Delta_{\pi}^{HT})^{0}$. It is
shown that the resummed result depends on the choice of the meson
wave functions used in calculation. Phenomenological effects of the
obtained results are discussed.
\end{abstract}
\pacs{12.38.-t, 13.60.Le, 13.87.Fh, 14.40.Aq,}

\keywords{higher twist, infrared renormalons, pion wave function}
\maketitle

\section{\bf Introduction}
The large-order behavior of a perturbative expansion in gauge
theories is inevitably dominated by the factorial growth of
renormalon diagrams [1-4]. In the case of quantum chromodynamics
(QCD), the coefficients of perturbative expansions in the QCD
coupling $\alpha_{s}$ can increase dramatically even at low orders.
This fact, together with the apparent freedom in the choice of
renormalization scheme and renormalization scales, limits the
predictive power of perturbative calculations, even in applications
involving large momentum transfers, where $\alpha_{s}$ is
effectively small.

A number of theoretical approaches have been developed to reorganize
the perturbative expansions in a effort to improve the
predictability of the perturbative QCD (pQCD). For example,
optimized scale and scheme choices have been proposed, such as the
method of effective charges (ECH) [5], the principle of minimal
sensitivity (PMS) [6], and the Brodsky-Lepage- Mackenize (BLM)
scale-setting prescription [7] and its generalizations [8-20]. In
[4] developments include the resummation of the formally divergent
renormalon series and the parametrization of related higher-twist
power-suppressed contributions.

In general, a factorially divergent renormalon series arises when
one integrates over the logarithmically running coupling
$\alpha_{s}(k^2)$ in a loop diagram. Such contributions do not occur
in conformally invariant theories which have a constant coupling. Of
course, in the physical theory, the QCD coupling does run.

Among the fundamental predictions of QCD are asymptotic scaling laws
for large-angle exclusive processes [21-27]. QCD counting rules were
formalized in Refs.[22,23]. These reactions probe hadronic
constituents at large relative momenta, or equivalently, the
hadronic wave function at short distances. Important examples of
exclusive amplitudes are provided by the electromagnetic form
factors of mesons. Since there is little direct evidence with which
to compare the predictions, it is fortunate that short-distance wave
functions also control a wide variety of processes at large
transverse momentum. In particular, the meson wave function
determines the leading higher-twist contribution to meson production
at high $p_T$.

The hadronic wave functions in terms of quark and gluon degrees of
freedom play an important role in the quantum chromodynamics
predictions for hadronic processes. In the perturbative  QCD theory,
the hadronic distribution amplitudes and structure functions which
enter exclusive and inclusive processes via the factorization
theorems at high momentum transfers can be determined by the
hadronic wave functions, and therefore they are the underlying links
between hadronic phenomena in QCD at large distances
(nonperturbative) and small distances (perturbative). If the
hadronic wave functions were accurately known, then we could
calculate the hadronic distribution amplitude and structure
functions for exclusive and inclusive processes in QCD. Conversely,
these processes also can provide phenomenological constraints on the
hadronic distribution amplitudes, the hadronic structure functions,
and thereby the hadronic wave functions. From another point of view,
as different wave functions may give the same distribution
amplitude, there are still ambiguities about the wave function even
if we know the exact form of the distribution amplitude.

In principle the hadronic wave functions determine all properties of
hadrons. From the relation between the wave function and measurable
quantities we can get some constraints on the general properties of
the hadronic wave functions. Note that only the lowest
$q\overline{q}$ Fock state contributes to the leading scaling
behavior; other Fock-state contributions are suppressed by powers of
$1/Q^2$.

The frozen coupling constant approach can be applied for
investigation, not only exclusive processes, but also for the
calculation of higher-twist contributions to some inclusive
processes, for example as large -$p_{T}$ meson photoproduction [28],
two-jet+meson production in the electron-positron annihilation [29].
In the works [28,29] for calculation of integrals, such as
\begin{equation}
I\sim \int\frac{\alpha_{s}(\hat {Q}^2)\Phi(x,\hat{Q}^2)}{1-x}dx
\end{equation}
the frozen coupling constant approach was used. According to Ref.[7]
should be noted that in pQCD calculations the argument of the
running coupling constant in both,  the renormalization and
factorization scale $\hat{Q}^2$ should be taken equal to the square
of the momentum transfer of a hard gluon in a corresponding Feynman
diagram. But defined in this way, $\alpha_{s}(\hat{Q}^2)$ suffers
from infrared singularities. For example in our work [30],
$\hat{Q}^2$ equals to $(x_{1}-1)\hat{u}$ and $-x_{1}\hat{t}$, where
$\hat{u}$, $\hat{t}$ are the subprocess's Mandelstam invariants.
Therefore, in the soft regions $x_{1}\rightarrow 0$,
$x_{2}\rightarrow 0$ integrals (1.1) diverge and for their
calculation some regularization methods of $\alpha_{s}(Q^2)$ in
these regions are needed. The power-suppressed corrections arising
from the soft end-point regions to the single meson photoproduction
process were computed in[31]. In Ref. [31], the evolution of the
meson wave function on the factorization scale was ignored. In the
present work, we take into account this evolution as well. In
Ref.[32],the authors investigated the phenomenology of infrared
renormalons in inclusive processes. The dispersive approach has been
devised to extend properly modified perturbation theory calculations
towards the low-energy region [33]. Connections between power
corrections for the three Deep Inelastic Scattering sum rules have
also been explored in [34].

Investigation of the infrared renormalon effects in various
inclusive and exclusive processes is one of the most important and
interesting problems in the perturbative QCD. It is known that
infrared renormalons are responsible for factorial growth of
coefficients in perturbative series for the physical quantities.
But, these divergent  series can be resummed by means of the Borel
transformation [1] and the principal  value prescription [35], and
effects of infrared renormalons can be taken into account by a
scale-setting procedure
$\alpha_{s}(Q^2)\rightarrow\alpha_{s}(exp(f(Q^2))Q^2)$ at the
one-loop order results. Technically, all-order resummation of
infrared renormalons corresponds to the calculation of the one-loop
Feynman diagrams with the running coupling constant
$\alpha_{s}(-k^2)$ at the vertices or, alternatively, to calculation
of the same diagrams with nonzero gluon mass. Studies of infrared
renormalon problems have also opened new prospects for evaluation of
power-suppressed corrections to processes characteristics [36].
Power corrections can also be obtained by means of the Landau-pole
free expression for the QCD coupling constant. The most simple and
elaborated variant of the dispersive approach, the Shirkov and
Solovtsov analytic perturbation theory, was formulated in Ref.[37].

By taking these points into account, it may be argued that the
analysis of the higher-twist effects on the dependence of the pion
wave function in pion production at proton-proton collisions by the
running coupling (RC) approach[38], are significant from both
theoretical and experimental points of view.

We will show that higher-twist terms contribute substantially to the
inclusive meson cross section at moderate transverse momenta. In
addition, we shall demonstrate that higher-twist reactions
necessarily dominate in the kinematic limit where the transverse
momentum approaches the phase-spase boundary.

A precise measurement of the inclusive charged pion production cross
section at $\sqrt s=62.4\,\,GeV$ and $\sqrt s=200\,\,GeV$ is
important for the proton-proton collisions program at  the
Relativistic Heavy Ion Collider (RHIC) at the Brookhaven National
Laboratory. The result of our calculations are based on the
proton-proton collisions at $\sqrt s=62.4\,\,GeV$ and $\sqrt
s=200\,\,GeV$

Another important aspect of this study is the choice of the meson
model wave functions. In this respect, the contribution of the
higher-twist Feynman diagrams to a pion production cross section in
proton-proton collisions has been computed by using various pion
wave functions. Also, higher-twist contributions which are
calculated by the running coupling constant and frozen coupling
constant approaches have been estimated and compared to each other.
Within this context, this paper is organized as follows: In
Sec.\ref{ht}, we provide formulas for the calculation of the
contribution of the high twist diagrams. In Sec. \ref{ir} we present
formulas and an analysis of the higher-twist effects on the
dependence of the pion wave function by the running coupling
constant approach. In Sec. \ref{lt}, we provide formulas for the
calculation of the contribution of the leading-twist diagrams. In
Sec. \ref{results}, we present the numerical results for the cross
section and discuss the dependence of the cross section on the pion
wave functions. We state our conclusions in Sec. \ref{conc}.

\section{CONTRIBUTION OF THE HIGH TWIST DIAGRAMS}
\label{ht} The higher-twist Feynman diagrams, which describe the
subprocess $q_1+\bar{q}_{2} \to \pi^{+}(\pi^{-})+\gamma$ for the
pion production in the proton-proton collision are shown in Fig.1.
In the higher-twist diagrams, the pion of a proton quark is directly
observed. Their $1/Q^2$ power suppression is caused by a hard gluon
exchange between pion constituents. The amplitude for this
subprocess can be found by means of the Brodsky-Lepage formula [26]

\begin{equation}
M(\hat s,\hat
t)=\int_{0}^{1}{dx_1}\int_{0}^{1}dx_2\delta(1-x_1-x_2)\Phi_{\pi}(x_1,x_2,Q^2)T_{H}(\hat
s,\hat t;x_1,x_2).
\end{equation}

In Eq.(2.1), $T_H$ is  the sum of the graphs contributing to the
hard-scattering part of the subprocess. The hard-scattering part for
the subprocess under consideration is $q_1+\bar{q}_{2} \to
(q_{1}\bar{q}_2)+\gamma$, where a quark and antiquark form a
pseudoscalar, color-singlet state $(q_1\bar{q}_2)$. Here
$\Phi(x_1,x_2,Q^2)$ is the pion wave function, i.e., the probability
amplitude for finding the valence $q_1\bar{q}_2$ Fock state in the
meson carry fractions $x_1$ and $x_2$, $x_1+x_2=1$. Remarkably, this
factorization is gauge invariant and only requires that the momentum
transfers in $T_H$ be large compared to the intrinsic mass scales of
QCD. Since the distribution amplitude and the hard-scattering
amplitude are defined without reference to the perturbation theory,
the factorization is valid to leading order in $1/Q$, independent of
the convergence of perturbative expansions.

The hard-scattering amplitude $T_H$ can be calculated in
perturbation theory and represented as a series in the QCD running
coupling constant $\alpha_s(Q^2)$. The function $\Phi$ is
intrinsically nonperturbative, but its evolution can be calculated
perturbatively.

The $q_{1}\overline{q}_{2}$ spin state used in computing $T_H$ may
be written in the form

\begin{equation}
\sum_{s_{1},s_{2}}
\frac{u_{s_1}({x}_{1}p_{M})\overline{v}_{s_{2}}({x}_{2}p_{M})}{\sqrt{x_1}
\sqrt{x_2}}\cdot N_{s_{1}s_{2}}^s=\left\{\begin{array}{ccc}
\frac{{\gamma}_{5}\hat {p}_{\pi}}{\sqrt{2}},\,\,\pi,\\\frac{\hat
{p}_{M}}{\sqrt{2}},\,\,\rho_L\,\,helicity \, 0,\\
\mp\frac{{\varepsilon}_{\mp}\hat {p}_{M}}{\sqrt{2}},\,\,
\rho{_T}\,\,helicity \pm1,\end{array}\right.
\end{equation}

where $\varepsilon_{\pm}=\mp(1/\sqrt{2})(0,1,\pm i,0)$ in a frame
with $(p_M)_{1,2}=0$ and the $N_{s_{1}s_{2}}^s$ project out a state
of a spin $s$, and $p_{M}$ is the four-momentum of the final meson.
In our calculation, we have neglected the pion and  proton masses.
Turning to extracting the contributions of the higher-twist
subprocesses, there are many kinds of leading-twist subprocesses in
$pp$ collisions as the background of the higher-twist subprocess
$q_1+q_2 \to \pi^{+}(or\,\, \pi^-)+\gamma$, such as $q+\bar{q} \to
\gamma+g(g \to \pi^{+}(\pi^{-}))$, $q+g \to \gamma+q(q \to
\pi^{+}(\pi^{-}))$, $\bar{q}+g \to \gamma+\bar{q}g(\bar{q} \to
\pi^{+}(\pi^{-}))$  etc. The contributions from these leading-twist
subprocesses strongly depend on some phenomenological factors, for
example, quark and gluon distribution functions in the proton and
fragmentation functions of various constituents, \emph{etc}. Most of
these factors have not been well determined, neither theoretically
nor experimentally. Thus they cause very large uncertainty in the
computation of the cross section of process $pp \to \pi^{+}(or\,\,
\pi^{-})+\gamma +X$. In general, the magnitude of this uncertainty
is much larger than the sum of all the higher-twist contributions,
so it is very difficult to extract the higher-twist contributions.

The Mandelstam invariant variables for subprocesses $q_1+\bar{q}_{2} \to \pi^{+}(\pi^{-})+\gamma$ are defined as
\begin{equation}
\hat s=(p_1+p_2)^2,\quad \hat t=(p_1-p_{\pi})^2,\quad \hat
u=(p_1-p_{\gamma})^2.
\end{equation}

In our calculation, we have also neglected the quark masses. We have
aimed to calculate the pion production cross section and to fix the
differences due to the use of various pion model functions. We have
used five different wave functions: the asymptotic wave function
(ASY), the Chernyak-Zhitnitsky wave function [27,39], the wave
function in which two nontrivial Gegenbauer coefficients $a_2$ and
$a_4$ have been extracted from the CLEO data on the
$\gamma\gamma^{\star} \to \pi^0$ transition form factor [40], the
Braun-Filyanov pion wave functions [41] and  the Bakulev-Mikhailov-
Stefanis pion wave function[42]. The wave functions of pions also
are developed in Refs.[43-45] by the Dubna group. In Ref.[40], the
authors have used the QCD light-cone sum rules approach and have
included into their analysis the NLO perturbative and twist-four
corrections. They found that in the model with two nonasymptotic
terms, at the scale $\mu_0=2.4 \,\,GeV$, $a_2=0.19$,\,$a_4=-0.14$.
$$
\Phi_{asy}(x)=\sqrt{3}f_{\pi}x(1-x),\quad
\Phi_{CZ}(x,\mu_{0}^2)=5\Phi_{asy}(2x-1)^2,\\
$$
$$
\Phi_{CLEO}(x,\mu_{0}^2)=\Phi_{asy}(x)[1+0.405(5(2x-1)^2-1)-0.4125((2x-1)^4-14(2x-1)^2+1),
$$
$$
\Phi_{BF}(x,\mu_{0}^2)=\Phi_{asy}(x)[1+0.66(5(2x-1)^2-1)+0.4687((2x-1)^4-14(2x-1)^2+1)],
$$
\begin{equation}
\Phi_{BMS}(x,\mu_{0}^2)=\Phi_{asy}(x)[1+0.282(5(2x-1)^2-1)-0.244((2x-1)^4-14(2x-1)^2+1)],
\end{equation}

where $f_{\pi}=0.923 GeV$ is the pion decay constant. Here, we have
denoted by $x\equiv x_1$, the longitudinal fractional momentum
carried by the quark within the meson. Then, $x_2=1-x$ and
$x_1-x_2=2x-1$. The pion wave function is symmetric under the
replacement $x_1-x_2\leftrightarrow x_2-x_1$. The model functions
can be written as
$$
\Phi_{asy}(x)=\sqrt{3}f_{\pi}x(1-x),
$$
$$
\Phi_{CZ}(x,\mu_{0}^2)=\Phi_{asy}(x)\left[C_{0}^{3/2}(2x-1)+\frac{2}{3}C_{2}^{3/2}(2x-1)\right],
$$
$$
\Phi_{CLEO}(x,\mu_{0}^2)=\Phi_{asy}(x)\left[C_{0}^{3/2}(2x-1)+0.27C_{2}^{3/2}(2x-1)-0.22C_{4}^{3/2}(2x-1)\right],
$$
$$
\Phi_{BF}(x,\mu_{0}^2)=\Phi_{asy}(x)\left[C_{0}^{3/2}(2x-1)+0.44C_{2}^{3/2}(2x-1)+
0.25C_{4}^{3/2}(2x-1)\right],\\
$$
\begin{equation}
\Phi_{BMS}(x,\mu_{0}^2)=\Phi_{asy}(x)\left[C_{0}^{3/2}(2x-1)+0.188C_{2}^{3/2}(2x-1)-0.13C_{4}^{3/2}(2x-1)\right],
\end{equation}
$$
C_{0}^{3/2}(2x-1)=1,\,\,C_{2}^{3/2}(2x-1)=\frac{3}{2}(5(2x-1)^2-1),
$$
$$
C_{4}^{3/2}(2x-1)=\frac{15}{8}(21(2x-1)^4-14(2x-1)^2+1).
$$

Several important nonperturbative tools have been developed which
allow specific predictions for the hadronic wave functions directly
from theory and experiments. The QCD sum-rule technique and lattice
gauge theory provide constraints on the moments of the hadronic
distribution amplitude. As is seen from Eq.(2.5) wave functions of
meson, which were constructed from theory and experiment strongly
depend on the methods is applied. However, the correct pion wave
function is still an open problem in QCD. It is known that the pion
wave function can be expanded over the eigenfunctions of the
one-loop Brodsky-Lepage equation, \emph{i.e.}, in terms of the
Gegenbauer polynomials $\{C_{n}^{3/2}(2x-1)\},$

\begin{equation}
\Phi_{\pi}(x,Q^2)=\Phi_{asy}(x)\left[1+\sum_{n=2,4..}^{\infty}a_{n}(Q^2)C_{n}^{3/2}(2x-1)\right],
\end{equation}

The evolution of the wave function on the factorization scale $Q^2$
is governed by the functions $a_n(Q^2)$,
\begin {equation}
a_n(Q^2)=a_n(\mu_{0}^2)\left[\frac{\alpha_{s}(Q^2)}{\alpha_{s}(\mu_{0}^2)}\right]^{\gamma_n/\beta_0},
\end{equation}
$$
\frac{\gamma_2}{\beta_{0}}=\frac{50}{81},\,\,\,\frac{\gamma_4}{\beta_{0}}=\frac{364}{405},\,\,
n_f=3.
$$

 In Eq.(2.7), $\{\gamma_n\}$ are anomalous dimensions defined by
the expression,

\begin{equation}
\gamma_n=C_F\left[1-\frac{2}{(n+1)(n+2)}+4\sum_{j=2}^{n+1}
\frac{1}{j}\right].
\end{equation}
The constants $a_n(\mu_{0}^2)=a_{n}^0$ are input parameters that
form the shape of the wave functions and which can be extracted from
experimental data or obtained from the nonperturbative QCD
computations at the normalization point $\mu_{0}^2$. The QCD
coupling constant $\alpha_{s}(Q^2)$ at the one-loop approximation is
given by the expression

\begin{equation}
\alpha_{s}(Q^2)=\frac{4\pi}{\beta_0 ln(Q^2/\Lambda^2)}.
\end{equation}
Here, $\Lambda$ is the fundamental QCD scale parameter, $\beta_0$ is
the QCD beta function one-loop coefficient, respectively,
$$
\beta_0=11-\frac{2}{3}n_f.
$$
The cross section for the higher-twist subprocess $q_1\bar{q}_{2}
\to \pi^{+}(\pi^{-})\gamma$ is given by the expression
\begin{equation}
\frac{d\sigma}{d\hat t}(\hat s,\hat t,\hat u)=\frac
{8\pi^2\alpha_{E} C_F}{27}\frac{\left[D(\hat t,\hat
u)\right]^2}{{\hat s}^3}\left[\frac{1}{{\hat u}^2}+\frac{1}{{\hat
t}^2}\right],
\end{equation}
where
\begin{equation}
D(\hat t,\hat u)=e_1\hat
t\int_{0}^{1}dx_1\left[\frac{\alpha_{s}(Q_1^2)\Phi_{\pi}(x_1,Q_1^2)}{1-x_1}\right]+e_2\hat
u\int_{0}^{1}dx_1\left[\frac{\alpha_{s}(Q_2^2)\Phi_{\pi}(x_1,Q_2^2)}{1-x_1}\right].
\end{equation}
Here $Q_{1}^2=(x_1-1)\hat u,\,\,\,\,$and $Q_{2}^2=-x_1\hat t$,\,\,
represent the momentum squared carried by the hard gluon in Fig.1,
$e_1(e_2)$ is the charge of $q_1(\overline{q}_2)$ and
$C_F=\frac{4}{3}$. The higher-twist contribution to the
large-$p_{T}$ pion production cross section in the process
$pp\to\pi^{+}(\pi^{-})+\gamma$ is [46]
\begin{equation}
\Sigma_{M}^{HT}\equiv E\frac{d\sigma}{d^3p}=\int_{0}^{1}\int_{0}^{1}
dx_1 dx_2 G_{{q_{1}}/{h_{1}}}(x_{1})
G_{{q_{2}}/{h_{2}}}(x_{2})\frac{\hat s}{\pi} \frac{d\sigma}{d\hat
t}(q\overline{q}\to \pi\gamma)\delta(\hat s+\hat t+\hat u).
\end{equation}
$$\pi E\frac{d\sigma}{d^3p}=\frac{d\sigma}{dydp_{T}^2},$$
$$\hat s=x_1x_2s,$$
$$\hat t=x_1t,$$
\begin{equation}
\hat u=x_2u,
\end{equation}
$$t= -m_T \sqrt{s} e^{-y}=-p_T \sqrt{s}e^{-y},$$
$$u= -m_T \sqrt {s} e^y=-p_T \sqrt{s}e^{y},$$
$$
x_1=-\frac{x_{2}u}{x_{2}s+t}=\frac{x_{2}p_{T}\sqrt s
e^{y}}{x_{2}s-p_{T}\sqrt s e^{-y}},
$$
$$
x_2=-\frac{x_{1}t}{x_{1}s+u}=\frac{x_{1}p_{T}\sqrt s
e^{-y}}{x_{1}s-p_{T}\sqrt s e^{y}},
$$

 where $m_T$ -- is the transverse mass of pion, which is given by
$$m_T^2=m^2+p_T^2.$$

Let us first consider the frozen coupling approach. In this approach
we take the four-momentum  square $\hat{Q}_{1,2}^2$ of the hard
gluon to be equal the pion's transverse momentum square
$\hat{Q}_{1,2}^2=p_{T}^2$. In this case, the QCD coupling constant
$\alpha_s$ in the integral (2.11) does not depend on the integration
variable. After this substitution calculation of integral (2.11)
becomes easy. Hence, the effective cross section obtained after
substitution of the integral (2.11) into the expression (2.10) is
referred as the frozen coupling effective cross section. We will
denote the higher-twist cross section obtained using the frozen
coupling constant approximation by $(\Sigma_{\pi}^{HT})^0$.

For a full discussion,  we consider a difference $\Delta_{\pi}^{HT}$
between the higher-twist cross section combinations
$\Sigma_{\pi^{+}}^{HT}$ and $\Sigma_{\pi^{-}}^{HT}$
\begin{equation}
\Delta_{\pi}^{HT}=\Sigma_{\pi^{+}}^{HT}
-\Sigma_{\pi^{-}}^{HT}=E_{{\pi}^{+}}\frac{d\sigma}{d^3p}
(pp\to\pi^{+}\gamma)-E_{{\pi}^{-}}\frac{d\sigma}{d^3p}(pp \to
\pi^{-}\gamma).
\end{equation}

We have extracted the following higher-twist subprocesses
contributing to the two covariant cross sections in Eq.(2.12)
\begin{equation}
\frac{{d\sigma}^1}{d\hat t}(u\bar{d} \to \pi^{+}\gamma),\,\,\,
\frac{{d\sigma}^2}{d\hat t}(\bar{d}u \to \pi^{+}\gamma),\,\,\,
\frac{{d\sigma}^3}{d\hat t}(\bar{u}d \to \pi^{-}\gamma),\,\,\,
\frac{{d\sigma}^4}{d\hat t}(d\bar{u} \to \pi^{-}\gamma),\,\,\,
\end{equation}
By charge conjugation invariance, we have
\begin {equation}
\frac{{d\sigma}^1}{d\hat t}(u\bar{d} \to
\pi^{+}\gamma)=\frac{{d\sigma}^3}{d\hat t}(\bar{u}d \to
\pi^{-}\gamma),\,\,and\,\,\, \frac{{d\sigma}^2}{d\hat t}(\bar{d}u
\to \pi^{+}\gamma)= \frac{{d\sigma}^4}{d\hat t}(d\bar{u} \to
\pi^{-}\gamma).
\end{equation}

\section{THE RUNNING COUPLING APPROACH AND HIGHER-TWIST MECHANISM}\label{ir}
In this section we shall calculate the integral (2.11) using the
running coupling constant method and also discuss the problem of
normalization of the higher-twist process cross section in the
context of the same approach.

As  is seen from (2.11), in general, one has to take into account
not only the dependence of $\alpha(\hat {Q}_{1,2}^2)$ on the scale
$\hat {Q}_{1,2}^2$, but also an evolution of $\Phi(x,\hat
{Q}_{1,2}^2)$ with $\hat {Q}_{1,2}^2$. The meson wave function
evolves in accordance with a Bethe-Salpeter-type equation.
Therefore, it is worth noting that, the renormalization scale
(argument of $\alpha_s$) should be equal to $Q_{1}^2=(x_1-1)\hat u$,
$Q_{2}^2=-x_1\hat t$, whereas the factorization scale [$Q^2$ in
$\Phi_{M}(x,Q^2)$] is taken independent from $x$, we take
$Q^2=p_{T}^2$. Such approximation does not considerably change the
numerical results, but the phenomenon considered in this article
(effect of infrared renormalons) becomes transparent. The main
problem in our investigation is the calculation of the integral in
(2.11) by the running coupling constant approach. The integral in
Eq.(2.11) in the framework of the running coupling approach takes
the form

\begin{equation}
I(\hat {Q}^2)=\int_{0}^{1}\frac{\alpha_{s}(\lambda
Q^2)\Phi_{M}(x,Q^2)dx}{1-x}.
\end{equation}

The $\alpha_{s}(\lambda Q^2)$ has the infrared singularity at
$x\rightarrow1$, if $\lambda=1-x$ and  as a result integral $(3.1)$
diverges (the pole associated with the denominator of the integrand
is fictitious, because $\Phi_{M}\sim(1-x)$, and therefore, the
singularity of the integrand at $x=1$  is caused only by
$\alpha_{s}((1-x)Q^2)$). For the regularization of the integral we
express the running coupling at scaling variable $\alpha_{s}(\lambda
Q^2)$ with the aid of the renormalization group equation in terms of
the fixed one $\alpha_{s}(Q^2)$. The renormalization group equation
for the running coupling $\alpha\equiv\alpha_{s}/\pi$ has the form
[35]

\begin{equation}
\frac{\partial\alpha(\lambda Q^2)}{\partial
ln\lambda}\simeq-\frac{\beta_{0}}{4}[\alpha(\lambda Q^2)]^2
\end{equation}
where
$$
\beta_{0}=11-\frac{2}{3}n_{f}.
$$
The solution of Eq.(3.2), with the initial condition
$$
\alpha(\lambda)|_{\lambda=1}=\alpha\equiv\alpha_{s}(Q^2)/\pi,
$$
is [35]

\begin{equation}
\frac{\alpha(\lambda)}{\alpha}=\left[1+\alpha\frac{\beta_{0}}{4}ln{\lambda}\right]^{-1}
\end{equation}

This transcendental equation can be solved iteratively by keeping
the leading $\alpha^kln^k\lambda$ order. This term is given by
\begin{equation}
\alpha(\lambda Q^2)\simeq\frac{\alpha_{s}}{1+ln\lambda/t}
\end{equation}

After substituting Eq.(3.4) into Eq.(2.11) we get
$$
D(Q^2)=e_{1}\hat t\int_{0}^{1}dx_{1}\frac{\alpha_{s}(\lambda
Q^2)\Phi_{M}(x,Q^2)}{1-x}+ e_{2}\hat
u\int_{0}^{1}dx_{1}\frac{\alpha_{s}(\lambda
Q^2)\Phi_{M}(x,Q^2)}{1-x}=
$$
$$
e_{1}\hat t\alpha_{s}(Q^2)\int_{0}^{1}dx_{1}
\frac{\Phi_{M}(x,Q^2)}{(1-x)(1+ln\lambda/t)}+ e_{2}\hat u
\alpha_{s}(Q^2)\int_{0}^{1}dx\frac{\Phi_{M}(x,Q^2)}{(1-x)(1+ln\lambda/t)}=
$$
$$
e_{1}\hat t\alpha_{s}(Q^2)\int_{0}^{1}dx
\frac{\Phi_{asy}(x)\left[1+\sum_{2,4,..}^{\infty}a_{n}(\mu_{0}^{2})
\left[\frac{\alpha_{s}(Q^2)}{\alpha_{s}(\mu_{0}^2)}\right]^{\gamma_{n}/\beta_{0}}C_{n}^{3/2}(2x-1)\right]}{(1-x)
(1+ln\lambda/t)}+
$$
\begin{equation}
e_{1}\hat u\alpha_{s}(Q^2)\int_{0}^{1}dx
\frac{\Phi_{asy}(x)\left[1+\sum_{2,4,..}^{\infty}a_{n}(\mu_{0}^{2})
\left[\frac{\alpha_{s}(Q^2)}{\alpha_{s}(\mu_{0}^2)}\right]^{\gamma_{n}/\beta_{0}}C_{n}^{3/2}(2x-1)\right]}{(1-x)(1+ln\lambda/t)},
\end{equation}

where $t=4\pi/\alpha_{s}(Q^2)\beta_{0}$

 The integral (3.5) is common and, of course, still divergent, but now it
is recast into a form, which is suitable for calculation. Using the
running coupling constant approach, this integral may be found as a
perturbative series in $\alpha_{s}(Q^2)$
\begin{equation}
D(Q^2)\sim
\sum_{n=1}^{\infty}\left(\frac{\alpha_{s}(Q^2)}{4\pi}\right)^nS_{n}.
\end{equation}
The expression coefficients $S_n$ can be written as power series in
the number of light quark flavors or, equivalently, as a series in
power of $\beta_0$.
$$
S_{n}=C_{n}\beta_{0}^{n-1}
$$
The coefficients $C_n$ of this series demonstrate factorial growth
$C_n\sim(n-1)!$, which might indicate an infrared renormalon nature
of divergences in the integral (3.5) and corresponding series (3.6).
The procedure for dealing with such ill-defined series is well
known; one has to perform the Borel transform of the series [15]
$$
B[D](u)=\sum_{n=0}^{\infty}\frac{D_n}{n!} u^n,
$$
then invert $B[D](u)$ to obtain the resummed expression (the Borel
sum) $D(Q^2)$. After this we can find directly the resummed
expression for $D(Q^2)$. The change of the variable $x$ to
$z=ln(1-x)$, as $ln(1-x)=ln\lambda$. Then,
\begin{equation}
D(Q^2)=e_{1}\hat t \alpha_{s}(Q^2) t \int_{0}^{1}
\frac{\Phi_{M}(x,Q^2)dx}{(1-x)(t+z)}+
e_{2}\hat u \alpha_{s}(Q^2) t \int_{0}^{1}
\frac{\Phi_{M}(x,Q^2)dx}{(1-x)(t+z)}
\end{equation}
For the calculation of the expression (3.7) we will apply the
inverse Laplase transform to Eq.(3.7) [47]. After this operation,
formula (3.7) is simplified and we can extract the Borel sum of the
perturbative series (3.6) and the corresponding Borel transform in
dependence from the wave functions of the meson, respectively. Also
after such manipulations the obtained expression can be used for
numerical computations.

 The inverse Laplace  transformation from $1/(t+z)$ has the form:
\begin{equation}
\frac{1}{t+z}=\int_{0}^{\infty}e^{-(t+z)u}du
\end{equation}
after inserting Eq.(3.8) into (3.7).
 Then, we obtain
$$
D(Q^2)=e_{1} \hat{t} \alpha_{s}(Q^2) t \int_{0}^{1}
\int_{0}^{\infty} \frac{\Phi_{M}(x,Q^2)e^{-(t+z)u}du dx}{(1-x)}+
$$
\begin{equation}
e_{2} \hat{u} \alpha_{s}(Q^2) t \int_{0}^{1} \int_{0}^{\infty}
\frac{\Phi_{M}(x,Q^2)e^{-(t+z)u}du dx}{(1-x)}.
\end{equation}
In the case of $\Phi_{asy}(x)$ for $D(Q^2)$, we get
\begin{equation}
D(Q^2)=\left(\frac{4\sqrt{3} \pi f_{\pi} e_{1} \hat
t}{\beta_{0}}+\frac{4\sqrt{3}\pi f_{\pi}e_{2}\hat u}{\beta_{0}}
\right)\left[\int_{0}^{\infty}due^{-tu}\left[\frac{1}{1-u}-
\frac{1}{2-u}\right]\right].
\end{equation}
In the case  of the $\Phi_{CZ}(x,Q^2)$ wave function, we find
$$
D(Q^2)=\left(\frac{4\sqrt{3}\pi f_{\pi}e_{1}\hat
t}{\beta_{0}}+\frac{4\sqrt{3}\pi f_{\pi}e_{2}\hat u}{\beta_{0}}
\right)\int_{0}^{\infty}du e^{-tu}
\left[\frac{1}{1-u}-\frac{1}{2-u}+\right.
$$
\begin{equation}
0.84\left[\frac{\alpha_{s}(Q^2)}{\alpha_{s}(\mu_{0}^2)}\right]^{50/81}
\left[\frac{4}{1-u}-\frac{24}{2-u}+\frac{40}{3-u}-
\left.\frac{20}{4-u}\right]\right],
\end{equation}
 In the case of the $\Phi_{CLEO}(x,Q^2)$ wave function, we get
$$
D(Q^2)=\left(\frac{4\sqrt{3}\pi f_{\pi}e_{1}\hat
t}{\beta_{0}}+\frac{4\sqrt{3}\pi f_{\pi}e_{2}\hat u}{\beta_{0}}
\right) \int_{0}^{\infty}du
e^{-tu}\left[\frac{1}{1-u}-\frac{1}{2-u}+ \right.
0.405\left[\frac{\alpha_{s}(Q^2)}{\alpha_{s}(\mu_{0}^2)}\right]^{50/81}\cdot
$$
$$
\left[\frac{4}{1-u}-\frac{24}{2-u}+\frac{40}{3-u}-
\frac{20}{4-u}\right]-0.4125\left[\frac{\alpha_{s}(Q^2)}{\alpha_{s}(\mu_{0}^2)}\right]^{364/405}\cdot
$$
\begin{equation}
\left.
\left[\frac{8}{1-u}-\frac{120}{2-u}+\frac{560}{3-u}-\frac{1112}{4-u}+\frac{1008}{5-u}-\frac{336}{6-u}\right]\right].
\end{equation}
Also, in the case of the $\Phi_{BMS}(x,Q^2)$ wave function, we get
$$
D(Q^2)=\left(\frac{4\sqrt{3}\pi f_{\pi}e_{1}\hat
t}{\beta_{0}}+\frac{4\sqrt{3}\pi f_{\pi}e_{2}\hat u}{\beta_{0}}
\right) \int_{0}^{\infty}du
e^{-tu}\left[\frac{1}{1-u}-\frac{1}{2-u}+ \right.
0.282\left[\frac{\alpha_{s}(Q^2)}{\alpha_{s}(\mu_{0}^2)}\right]^{50/81}\cdot
$$
$$
\left[\frac{4}{1-u}-\frac{24}{2-u}+\frac{40}{3-u}-
\frac{20}{4-u}\right]-0.244\left[\frac{\alpha_{s}(Q^2)}{\alpha_{s}(\mu_{0}^2)}\right]^{364/405}\cdot
$$
\begin{equation}
\left.
\left[\frac{8}{1-u}-\frac{120}{2-u}+\frac{560}{3-u}-\frac{1112}{4-u}+\frac{1008}{5-u}-\frac{336}{6-u}\right]\right].
\end{equation}
Equation(3.1) and (3.2) is nothing more than the Borel sum of the
perturbative series (3.6), and the corresponding Borel transform in
the case $\Phi_{asy}(x)$ is
\begin{equation}
B[D](u)=\frac{1}{1-u}-\frac{1}{2-u},
\end{equation}
in the case $\Phi_{CZ}(x,Q^2)$ is
\begin{equation}
B[D](u)=\frac{1}{1-u}-\frac{1}{2-u}+0.84\left(\frac{\alpha_{s}(Q^2)}{\alpha_{s}(\mu_{0}^2)}\right)^{50/81}
\left(\frac{4}{1-u}-\frac{24}{2-u}+\frac{40}{3-u}-\frac{20}{4-u}\right),
\end{equation}

 in the case $\Phi_{CLEO}(x,Q^2)$ is
$$
B[D](u)=\frac{1}{1-u}-\frac{1}{2-u}+0.405\left(\frac{\alpha_{s}(Q^2)}{\alpha_{s}(\mu_{0}^2)}\right)^{50/81}
\left(\frac{4}{1-u}-\frac{24}{2-u}+\frac{40}{3-u}-\frac{20}{4-u}\right)-
$$
\begin{equation}
0.4125\left(\frac{\alpha_{s}(Q^2)}{\alpha_{s}(\mu_{0}^2)}\right)^{364/405}\left(\frac{8}{1-u}-
\frac{120}{2-u}+\frac{560}{3-u}-\frac{1112}{4-u}+\frac{1008}{5-u}-\frac{336}{6-u}\right).
\end{equation}
and in the case $\Phi_{BMS}(x,Q^2)$ is
$$
B[D](u)=\frac{1}{1-u}-\frac{1}{2-u}+0.282\left(\frac{\alpha_{s}(Q^2)}{\alpha_{s}(\mu_{0}^2)}\right)^{50/81}
\left(\frac{4}{1-u}-\frac{24}{2-u}+\frac{40}{3-u}-\frac{20}{4-u}\right)-
$$
\begin{equation}
0.244\left(\frac{\alpha_{s}(Q^2)}{\alpha_{s}(\mu_{0}^2)}\right)^{364/405}\left(\frac{8}{1-u}-
\frac{120}{2-u}+\frac{560}{3-u}-\frac{1112}{4-u}+\frac{1008}{5-u}-\frac{336}{6-u}\right).
\end{equation}
The series (3.6) can be recovered by means of the following formula
$$
C_{n}=\left(\frac{d}{du}\right)^{n-1}B[D](u)\mid_{u=0}
$$
The Borel transform $B[D](u)$ has poles on the real $u$ axis at
$u=1;2;3;4;5;6,$ which confirms our conclusion concerning the
infrared renormalon nature of divergences in (3.6). To remove them
from Eqs.(3.11) and (3.12) some regularization methods have to be
applied. In this article we adopt the principal value prescription.
We obtain: in the case $\Phi_{asy}$
\begin{equation}
[D(Q^2)]^{res}=\left(\frac{4\sqrt{3}\pi f_{\pi}e_{1}\hat
t}{\beta_{0}}+\frac{4\sqrt{3}\pi f_{\pi}e_{2}\hat
u}{\beta_{0}}\right)\left[\frac{Li(\lambda)}{\lambda}-\frac{Li(\lambda^2)}{\lambda^2}\right],
\end{equation}
in the case $\Phi_{CZ}(x,Q^2)$
$$
[D(Q^2)]^{res}=\left(\frac{4\sqrt{3}\pi f_{\pi}e_{1}\hat
t}{\beta_{0}}+\frac{4\sqrt{3}\pi f_{\pi}e_{2}\hat
u}{\beta_{0}}\right)\left[\left[\frac{Li(\lambda)}{\lambda}-\frac{Li(\lambda^2)}{\lambda^2}\right]+
0.84\left(\frac{\alpha_{s}(Q^2)}{\alpha_{s}(\mu_{0}^2)}\right)^{50/81}\right.
$$
\begin{equation}
\left[4\frac{Li(\lambda)}{\lambda}-24\frac{Li(\lambda^2)}{\lambda^2}+
\left.40\frac{Li(\lambda^3)}{\lambda^3}-20\frac{Li(\lambda^4)}{\lambda^4}\right]\right],
\end{equation}
 in the case $\Phi_{CLEO}(x,Q^2)$
$$
[D(Q^2)]^{res}=\left(\frac{4\sqrt{3}\pi f_{\pi}e_{1}\hat
t}{\beta_{0}}+\frac{4\sqrt{3}\pi f_{\pi}e_{2}\hat
u}{\beta_{0}}\right)\left[\left(\frac{Li(\lambda)}{\lambda}-\frac{Li(\lambda^2)}{\lambda^2}\right)+\right.
0.405\left(\frac{\alpha_{s}(Q^2)}{\alpha_{s}(\mu_{0}^2)}\right)^{50/81}\left(4\frac{Li(\lambda)}{\lambda}-\right.
$$
$$
24\frac{Li(\lambda^2)}{\lambda^2}+40\frac{Li(\lambda^3)}{\lambda^3}-
\left.20\frac{Li(\lambda^4)}{\lambda^4}\right)-
0.4125\left(\frac{\alpha_{s}(Q^2)}{\alpha_{s}(\mu_{0}^2)}\right)^{364/405}\left(8\frac{Li(\lambda)}{\lambda}-\right.
120\frac{Li(\lambda^2)}{\lambda^2}+560\frac{Li(\lambda^3)}{\lambda^3}-
$$
\begin{equation}
1112\frac{Li(\lambda^4)}{\lambda^4}+
1008\frac{Li(\lambda^5)}{\lambda^5}-
\left.\left.336\frac{Li(\lambda^6)}{\lambda^6}\right)\right],
\end{equation}
also in the case $\Phi_{BMS}(x,Q^2)$
$$
[D(Q^2)]^{res}=\left(\frac{4\sqrt{3}\pi f_{\pi}e_{1}\hat
t}{\beta_{0}}+\frac{4\sqrt{3}\pi f_{\pi}e_{2}\hat
u}{\beta_{0}}\right)\left[\left(\frac{Li(\lambda)}{\lambda}-\frac{Li(\lambda^2)}{\lambda^2}\right)+\right.
0.282\left(\frac{\alpha_{s}(Q^2)}{\alpha_{s}(\mu_{0}^2)}\right)^{50/81}\left(4\frac{Li(\lambda)}{\lambda}-\right.
$$
$$
24\frac{Li(\lambda^2)}{\lambda^2}+40\frac{Li(\lambda^3)}{\lambda^3}-
\left.20\frac{Li(\lambda^4)}{\lambda^4}\right)-
0.244\left(\frac{\alpha_{s}(Q^2)}{\alpha_{s}(\mu_{0}^2)}\right)^{364/405}\left(8\frac{Li(\lambda)}{\lambda}-\right.
120\frac{Li(\lambda^2)}{\lambda^2}+560\frac{Li(\lambda^3)}{\lambda^3}-
$$
\begin{equation}
1112\frac{Li(\lambda^4)}{\lambda^4}+
1008\frac{Li(\lambda^5)}{\lambda^5}-
\left.\left.336\frac{Li(\lambda^6)}{\lambda^6}\right)\right],
\end{equation}
where $Li(\lambda)$ is the logarithmic integral for $\lambda>1$
defined as the principal value[48]
\begin{equation}
Li(\lambda)=P.V.\int_{0}^{\infty}\frac{dx}{lnx},\,\,\,
\lambda=Q^2/\Lambda^2.
\end{equation}

 Hence, the effective cross section obtained after substitution of the
expressions (3.10-3.12) into the expression (2.10) is referred as
the running coupling effective cross section. We will denote the
higher-twist cross section obtained using the running coupling
constant approach by $(\Sigma_{\pi}^{HT})^{res}$.

\section{CONTRIBUTION OF THE  LEADING-TWIST DIAGRAMS}\label{lt}
Regarding the higher-twist corrections to the pion production cross
section, a comparison of our results with leading-twist
contributions is crucial. We take two leading-twist subprocesses for
the pion production:(1) quark-antiquark annihilation $q\bar{q} \to
g\gamma$, in which the $\pi$ meson is indirectly emitted from the
gluon, $g \to \pi^{+}(\pi^{-})$ and (2) quark-gluon fusion, $qg \to
q\gamma $, with subsequent fragmentation of the final quark into a
meson, $q \to \pi^{+}(\pi^{-})$. The corresponding cross sections
are obtained in
\begin{equation}
\frac{d\sigma}{d\hat t}(q\bar{q} \to gq)=\frac{8}{9}\pi\alpha_E
\alpha_s(Q^2)\frac{e_{q}^2}{{\hat s}^2}\left(\frac{\hat t}{\hat
u}+\frac{\hat u}{\hat t}\right),
\end{equation}
\begin{equation}
\frac{d\sigma}{d\hat t}(qg \to q\gamma)=-\frac{\pi{e_{q}^2}\alpha_E
\alpha_s(Q^2)}{{3\hat s}^2}\left(\frac{\hat s}{\hat t}+\frac{\hat
t}{\hat s}\right).
\end{equation}
For the leading-twist contribution, we find
$$
\Sigma_{M}^{LT}\equiv E\frac{d\sigma}{d^3p}=\sum_{q}\int_{0}^{1}
dx_1 dx_2dz \left(G_{{q_{1}}/{h_{1}}}(x_{1})
G_{{q_{2}}/{h_{2}}}(x_{2})D_{g}^{\pi}(z)\frac{\hat s}{\pi
z^2}\frac{d\sigma}{d\hat t}(q\bar{q}\to g\gamma)+\right.
$$
\begin{equation}
\left.G_{{q_{1}}/{h_{1}}}(x_{1})
G_{{g}/{h_{2}}}(x_{2})D_{q}^{\pi}(z)\frac{\hat s}{\pi
z^2}\frac{d\sigma}{d\hat t}(qg \to q\gamma)\right) \delta(\hat
s+\hat t+\hat u),
\end{equation}
where
\begin{equation}
\hat s=x_{1}x_{2}s,\,\,\hat t=\frac{x_{1}t}{z},\,\,\hat
u=\frac{x_{2}u}{z},\,\, z=-\frac{x_{1}t+x_{2}u}{x_{1}x_{2}s}.
\end{equation}
$D_{g}^{\pi}(z)=D_{g}^{\pi^{+}}(z)=D_{g}^{\pi^{-}}(z)$ and
$D_{q}^{\pi}(z)$ represents gluon and quark  fragmentation functions
into a meson containing  gluon and quark of the same flavor. In the
leading-twist subprocess, the  $\pi$ meson is indirectly emitted
from the gluon and quark with the fractional momentum $z$. The
$\delta$ function can be expressed in terms of the parton kinematic
variables, and the $z$ integration can then be done. The final form
for the cross section is
$$
\Sigma_{M}^{LT}\equiv
E\frac{d\sigma}{d^3p}=\sum_{q}\int_{x_{1min}}^{1} dx_1
\int_{x_{2min}}^{1} dx_2 \left(G_{{q_{1}}/{h_{1}}}(x_{1})
G_{{q_{2}}/{h_{2}}}(x_{2})D_{g}^{\pi}(z) \cdot \frac{1}{\pi
z}\frac{d\sigma}{d\hat t}(q\bar{q} \to g\gamma)+\right.
$$
$$
\left.G_{{q_{1}}/{h_{1}}}(x_{1})
G_{{g}/{h_{2}}}(x_{2})D_{g}^{\pi}(z) \cdot \frac{1}{\pi
z}\frac{d\sigma}{d\hat t}(qg \to q\gamma)\right)=
$$
$$
\sum_{q}\int_{x_{1min}}^{1} dx_1 \int_{x_{2min}}^{1}
\frac{dx_2}{-(x_{1}t+x_{2}u)}\left(x_{1}G_{{q_{1}}/{h_{1}}}(x_{1})
sx_{2}G_{{q_{2}}/{h_{2}}}(x_{2})\frac{D_{g}^{\pi}(z)}{\pi}
\frac{d\sigma}{d\hat t}(q\overline{q} \to g\gamma)+\right.
$$
\begin{equation}
\left.x_{1}G_{{q_{1}}/{h_{1}}}(x_{1})
sx_{2}G_{{g}/{h_{2}}}(x_{2})\frac{D_{g}^{\pi}(z)}{\pi}\frac{d\sigma}{d\hat
t}(qg \to q\gamma)\right).
\end{equation}

\section{NUMERICAL RESULTS AND DISCUSSION}\label{results}

In this section, we discuss the numerical results for higher-twist
effects with higher-twist contributions calculated in the context of
the running coupling constant  and frozen coupling approaches on the
dependence of the chosen meson wave functions in the process $pp \to
\pi^{+}(or\,\, \pi^{-})\gamma$. In the calculations, we use the
asymptotic  wave function $\Phi_{asy}$, the Chernyak-Zhitnitsky
$\Phi_{CZ}$, the pion wave function (from which two nontrivial
Gegenbauer coefficients $a_2$ and $a_4$ have been extracted from the
CLEO data on the $\pi^{0}\gamma$ transition form factor[40]), the
Braun-Filyanov pion wave functions [41], and the Bakulev-Mikhailov-
Stefanis pion wave function[ 42]. In Ref.[40], the authors have used
the QCD light-cone sum rules approach and included into their
analysis the NLO perturbative and twist-four corrections. For the
higher-twist subprocess, we take $q_1+\bar{q}_{2} \to
(q_1\bar{q}_2)+\gamma$ and we have extracted the following four
higher-twist subprocesses  contributing to $pp\to
\pi^{+}(or\,\,\pi^{-})\gamma$ cross sections: $u\bar{d} \to
\pi^{+}\gamma,$ $\bar{d}u \to \pi^{+}\gamma$, $\bar{u}d \to
\pi^{-}\gamma$, $d\bar{u} \to \pi^{-}\gamma$ contributing to cross
sections. For the dominant leading-twist subprocess for the pion
production, we take the quark-antiquark annihilation $q\bar{q} \to
g\gamma$, in which the $\pi$ meson is indirectly emitted from the
gluon and quark-gluon fusion, $qg \to q\gamma $, with subsequent
fragmentation of the final quark into a meson, $q \to
\pi^{+}(\pi^{-})$.  As an example for the quark distribution
function inside the proton, the MRST2003c package [49] has been
used. The higher twist subprocesses probe the meson wave functions
over a large range of $Q^2$ squared momentum transfer, carried by
the gluon. Therefore, in the diagram given in Fig.1 we take
$Q_{1}^2=(x_1-1){\hat u}$, $Q_{2}^2=-x_1\hat t$ , which we have
obtained directly from the higher-twist subprocesses diagrams. The
same $Q^2$ has been used as an argument of $\alpha_s(Q^2)$ in the
calculation of each diagram.

The results of our numerical calculations are plotted in Figs.2-31.
First of all, it is very interesting to compare the resummed higher-
twist cross sections with the ones obtained in the framework of the
frozen coupling approach. In Figs.2-4 we show the dependence of
higher-twist cross sections $(\Sigma_{\pi^{+}}^{HT})^{0}$ calculated
in the context of the frozen coupling,
$(\Sigma_{\pi^{+}}^{HT})^{res}$ in the context of the running
coupling constant approaches and also the ratio
$R=(\Sigma_{\pi^{+}}^{HT})^{res}$/$\Sigma_{\pi^{+}}^{HT})^{0}$ as a
function of the pion transverse momentum $p_{T}$ for different pion
wave functions at $y=0$. It is seen that the values of cross
sections $(\Sigma_{\pi^{+}}^{HT})^{0}$,
$(\Sigma_{\pi^{+}}^{HT})^{res}$, and $R$ for fixed $y$ and $\sqrt s$
depend on the choice of the pion wave function. As seen from
Figs.2-3 in both cases, frozen coupling and running coupling
constant approaches the higher-twist differential cross section is
monotically decreasing with an increase in the transverse momentum
of the pion. In Figs.5 and 6, we shows the dependence of the ratio
$(\Sigma_{\pi^{+}}^{HT})^{0}$/$(\Sigma_{\pi^{+}}^{LT})$ and
$(\Sigma_{\pi^{+}}^{HT})^{res}$/$(\Sigma_{\pi^{+}}^{LT})$ as a
function of the pion transverse momentum $p_{T}$ for different pion
wave functions. Here $(\Sigma_{\pi^{+}}^{HT})^{res}$,
$(\Sigma_{\pi^{+}}^{HT})^0$ are the higher-twist cross sections
calculated in the context of the running coupling method and in the
framework of the frozen coupling approach and
$(\Sigma_{\pi^{+}}^{LT})$ is the leading-twist cross section,
respectively. As seen from Fig.6, in the region
$2\,\,GeV/c<p_T<5\,\,GeV/c$  higher-twist cross section calculated
in the context of the running coupling method is suppressed by about
2 orders of magnitude relative to the leading-twist cross section,
but in the region $5\,\,GeV/c<p_T\leq 30\,\,GeV/c$ is comparable
with the cross section of leading-twist. In Figs.7-10 we show the
dependence $(\Delta_{\pi}^{HT})^{0}$, $(\Delta_{\pi}^{HT})^{res}$,
the ratio $r$=($\Delta_{\pi}^{HT})^{res}$/$(\Delta_{\pi}^{HT})^{0}$,
and the ratio ($\Delta_{\pi}^{HT})^{res}$/$(\Delta_{\pi}^{LT})$ as a
function of the pion transverse momentum $p_{T}$ for the pion wave
functions. Here, $(\Delta_{\pi}^{HT})^0=(\Sigma_{\pi^{+}}^{HT})^0-
(\Sigma_{\pi^{-}}^{HT})^0$ and
$(\Delta_{\pi}^{HT})^{res}=(\Sigma_{\pi^{+}}^{HT})^{res}-
(\Sigma_{\pi^{-}}^{HT})^{res}$. As seen from Figs.7 and 8, the
higher-twist differential cross section is decreasing with an
increase in the transverse momentum of the pion. As is seen from
Fig.9, when the transverse momentum of the pion is increasing, the
ratio $r$  is decreasing. But, as shown in Fig.9, in the region
$2\,\,GeV/c<p_T<25\,\,GeV/c$ higher-twist cross section calculated
in the context of the running coupling method is suppressed by about
3 orders of magnitude relative to the higher-twist cross section
calculated in the framework of the frozen coupling method. The
dependence, as shown in Fig.10, is identically equivalent to Fig.6.

In Figs.11-16, we have depicted higher-twist cross sections, ratios
$(\Sigma_{\pi^{+}}^{HT})^{0}$, $(\Sigma_{\pi^{+}}^{HT})^{res}$,
$R=(\Sigma_{\pi^{+}}^{HT})^{res}$/$(\Sigma_{\pi^{+}}^{HT})^{0}$,
$r$=($\Delta_{\pi}^{HT})^{res}$/$(\Delta_{\pi}^{HT})^{0}$,
$(\Delta_{\pi}^{HT})^{0}$/$(\Delta_{\pi}^{LT})$, and
$(\Delta_{\pi}^{HT})^{res}$/$(\Delta_{\pi}^{LT})$ as a function of
the rapidity $y$ of the pion at $\sqrt s=62.4\,\,GeV$ and
$p_T=4.9\,\,GeV/c$. At $\sqrt s=62.4\,\,GeV$ and $p_T=4.9\,\,GeV/c$,
the pion rapidity lies in the region $-2.52\leq y\leq2.52$.

As seen from Figs.13-14, in the region ($-2.52\leq y\leq -1.92$),
the ratio for all wave functions increase with an increase of the
$y$ rapidity of the pion and has a maximum approximately at the
point $y=-1.92$. Besides that, the ratio decreases with an increase
in the $y$ rapidity of the pion. As is seen from Figs.13-14, the
ratios $R$ and $r$ are very sensitive to the choice of the meson
wave functions. But, as seen from Fig.15, the ratio
$(\Sigma_{\pi^{+}}^{HT})^{0}$/$(\Sigma_{\pi^{+}}^{LT})$ for all wave
functions has a minimum approximately at the point $y=-1.92$. In
Fig.16 we show the ratio
$(\Delta_{\pi}^{HT})^{res}$/$(\Delta_{\pi}^{LT})$ as a function of
the rapidity $y$ of the pion. As seen from Fig.16, with an increase
of the $y$ rapidity of the pion  the ratio increases. It should be
noted that the magnitude of the higher-twist cross section for the
pion wave function $\Phi_{BMS}(x,Q^2)$ is very close to the
asymptotic wave function$\Phi_{asy}(x)$. The higher-twist
corrections and ratio are very sensitive to the choice of the pion
wave function. Also, the distinction between $R(\Phi_{asy}(x))$ with
$R(\Phi_{CLEO}(x,Q^2))$, $R(\Phi_{CZ}(x,Q^2))$,
$R(\Phi_{BF}(x,Q^2))$   and $R(\Phi_{BMS}(x,Q^2))$ have been
calculated. For example, in the case of $\sqrt s=62.4\,\,GeV$,
$y=0$, the distinction between $R(\Phi_{asy}(x))$ with
$R(\Phi_{i}(x,Q^2))$\,\,(i=CLEO, CZ, BF, BMS) as a function of the
pion transverse momentum $p_{T}$ is shown in Table \ref{table1}.
Thus, the distinction between $R(\Phi_{asy}(x))$ and
$R(\Phi_{i}(x,Q^2))(i=CLEO,CZ,BF)$ is maximum at $p_T=20\,\,GeV/c$,
with $R(\Phi_{BMS}(x))$  at  $p_T=2\,\,GeV/c$ but the distinction
between $R(\Phi_{asy}(x))$ with $R(\Phi_{i}(x,Q^2))(i=CLEO, CZ,BF)$
is minimum at $p_T=2\,\,GeV/c$, with $R(\Phi_{BMS}(x))$  at
$p_T=20\,\,GeV/c$  and increase with an increase in $p_T$. Such a
behavior  of $R$ may be explained by reducing all moments of the
pion model wave functions to those of $\Phi_{asy}(x)$ for high
$Q^2$. Also, we have calculated the distinction between
$R(\Phi_{asy}(x))$ with $R(\Phi_{CLEO}(x,Q^2))$,
$R(\Phi_{CZ}(x,Q^2))$, $R(\Phi_{BF}(x,Q^2))$ and
$R(\Phi_{BMS}(x,Q^2))$ as a function of the rapidity $y$ of the
pion. For example, in the case of $\sqrt s=62.4GeV$,
$p_{T}=4.9GeV/c$  the distinction between $R(\Phi_{asy}(x))$ with
$R(\Phi_{i}(x,Q^2))$\,\,(i=CLEO, CZ, BF, BMS) as a function of the
rapidity $y$ of the pion is presented in Table \ref{table2}

We have also carried out comparative calculations in the
center-of-mass energy $\sqrt s=200\,\,GeV$. The results of our
numerical calculations in the center-of-mass energies $\sqrt
s=200\,\,GeV$ are plotted in Figs.17-31. Analysis of our
calculations at the center-of-mass energies $\sqrt s=62.4\,\,GeV$
and $\sqrt s=200\,\,GeV$, show that with the increase in beam energy
values of the cross sections, ratio
$R=(\Sigma_{\pi^{+}}^{HT})^{res}/(\Sigma_{\pi^{+}}^{HT})^{0}$, and
contributions of higher-twist to the cross section decrease by about
1-3 order. Therefore the experimental investigation  of higher-twist
effects include renormalon effects conveniently in low energy.
 On the other hand, the higher-twist corrections and ratios $R$ and
$r$ are very sensitive to the choice of the pion wave function.
Also, the distinction between $R(\Phi_{asy}(x))$ with
$R(\Phi_{CLEO}(x,Q^2))$, $R(\Phi_{CZ}(x,Q^2))$,
$R(\Phi_{BF}(x,Q^2))$  and $R(\Phi_{BMS}(x,Q^2))$ have been
calculated. For example, in the case of $\sqrt s=200\,\,GeV$, $y=0$,
the distinction between $R(\Phi_{asy}(x))$ with
$R(\Phi_{i}(x,Q^2))$\,\,(i=CLEO, CZ, BF, BMS) as a function of the
pion transverse momentum $p_{T}$ is shown in Table \ref{table3}.
Thus, the distinction between $R(\Phi_{asy}(x))$ with
$R(\Phi_{i}(x,Q^2))$,\,\,(i=CZ, CLEO, BF,) is maximum at
$p_T=35\,\,GeV/c$, with $R(\Phi_{BMS}(x))$  at  $p_T=10\,\,GeV/c$,
but the distinction between $R(\Phi_{asy}(x))$ with
$R(\Phi_{CZ}(x,Q^2))$, $R(\Phi_{CLEO}(x,Q^2))$,
$R(\Phi_{BF}(x,Q^2))$ is minimum at $p_T=10\,\,GeV/c$, with
$R(\Phi_{BMS}(x))$ at $p_T=95\,\,GeV/c$  and increase with an
increase in $p_T$. Also, we have calculated the distinction between
$R(\Phi_{asy}(x))$ with $R(\Phi_{CLEO}(x,Q^2))$,
$R(\Phi_{CZ}(x,Q^2))$, $R(\Phi_{BF}(x,Q^2))$ and
$R(\Phi_{BMS}(x,Q^2))$as a function of the rapidity $y$ of the pion.
For example, in the case of $\sqrt s=200GeV$, $p_{T}=15.5GeV/c$  the
distinction between $R(\Phi_{asy}(x))$ with $R(\Phi_{i}(x,Q^2))$ as
a function of the rapidity $y$ of the pion is presented in Table
\ref{table4}. The calculations show that the ratio
$R(\Phi_{i}(x,Q^2))$/$R(\Phi_{asy}(x))$, (i=CLEO, CZ, BF, BMS) for
all values of the transverse momentum $p_T$ of the pion identically
equivalent to ratio $r(\Phi_{i}(x,Q^2))$/$r(\Phi_{asy}(x))$.

\section{Concluding Remarks}\label{conc}
In this work we have calculated the single meson inclusive
production via higher-twist mechanism and obtained the expressions
for the subprocess $q\overline{q}\rightarrow M \gamma$ cross section
for mesons with symmetric wave functions. For calculation of the
cross section we have applied the running coupling constant method
and revealed infrared renormalon poles in the cross section
expression. Infrared renormalon induced divergences have been
regularized by means of the principal value prescripton and the
resummed expression (the Borel sum) for the higher-twist cross
section has been found. The higher-twist cross sections were
calculated in the frozen coupling and running coupling approaches.
The resummed higher-twist cross section differs from that found
using the frozen coupling approach, in some regions, considerably.
Also we demonstrated that higher-twist contributions to single meson
production cross section in the proton -proton collisions have
important phenomenological consequences. We have obtained very
interesting results. The ratio $R$ for all values of the transverse
momentum $p_{T}$ and of the rapidity of the pion identically
equivalent to ratio $r$. Our investigation enables us to conclude
that the higher-twist pion production cross section in the
proton-proton collisions depends on the form of the pion model wave
functions and may be used for their study. Analysis of our
calculations shows that the magnitude of cross sections of the
leading-twist is larger than the higher-twist cross sections ones
calculated in the frozen coupling approach in 2-4 order. But, in
some regions of transverse momentum of the pion, the higher-twist
cross section calculated in the context of the running coupling
method is comparable with the cross sections of leading-twist.
Further investigations are needed in order to clarify the role of
high twist effects  in this process. We have demonstrated that the
resummed result depends on the pion model wave functions used in
calculations. The proton-proton collisions provide us with a new
opportunity to probe a proton's internal structure. In particular,
meson production in proton-proton collisions  takes into account
infrared renormalon effects: this opens a window  toward new types
of parton distributions-- chiral-odd distributions
$h_{1}(x,\mu^{2})$ and $h_{L}(x,\mu^{2})$ which can not be measured
by the deep inelastic lepton-proton scatterings.

\newpage

\begin{table}[ht]
\begin{center}
\begin{tabular}{|c|c|c|c|c|c} \hline
$p_{T},GeV/c$ & $\frac{R(\Phi_{CLEO}(x,Q^2))}{R(\Phi_{asy}(x))}$ &
$\frac{R(\Phi_{CZ}(x,Q^2))}{R(\Phi_{asy}(x))}$ &
$\frac{R(\Phi_{BF}(x,Q^2))}{R(\Phi_{asy}(x))}$ &
$\frac{R(\Phi_{BMS}(x,Q^2))}{R(\Phi_{asy}(x))}$ \\
\hline
  2 & 0.557  & 0.299 & 0.462 & 9.813 \\ \hline
  6 & 1.744 & 0.513 & 1.5 &2.208 \\ \hline
  20 & 7.273  & 6.065 & 6.311& 3.357\\ \hline

\end{tabular}
\end{center}
\caption{The distinction between $R(\Phi_{asy}(x))$ with
$R(\Phi_{i}(x,Q^{2}))$  (i=CLEO, CZ, BF, BMS) at c.m.
 energy $\sqrt s=62.4\,\,GeV$.} \label{table1}
\end{table}

\begin{table}[ht]
\begin{center}
\begin{tabular}{|c|c|c|c|c|c} \hline
$y$ & $\frac{R(\Phi_{CLEO}(x,Q^2))}{R(\Phi_{asy}(x))}$ &
$\frac{R(\Phi_{CZ}(x,Q^2))}{R(\Phi_{asy}(x))}$ &
$\frac{R(\Phi_{BF}(x,Q^2))}{R(\Phi_{asy}(x))}$&
$\frac{R(\Phi_{BMS}(x,Q^2))}{R(\Phi_{asy}(x))}$ \\
\hline -2.52& 11.285 & 2.094 & 10.162 & 3.232 \\ \hline -1.92 & 0.36
&0.279
& 0.446&5.103 \\ \hline 0.78& 0.076  & 0.945 & 6.213 & 2.392\\
\hline
\end{tabular}
\end{center}
\caption{The distinction between $R(\Phi_{asy}(x))$ with
$R(\Phi_{i}(x,Q^{2}))$ (i=CLEO, CZ, BF, BMS) at c.m. energy $\sqrt
s=62.4\,\,GeV$ and $p_T=4.9\,\, GeV/c$.} \label{table2}
\end{table}

\begin{table}[ht]
\begin{center}
\begin{tabular}{|c|c|c|c|c|c}                 \hline
$p_{T},GeV/c$ & $\frac{R(\Phi_{CLEO}(x,Q^2))}{R(\Phi_{asy}(x))}$ &
$\frac{R(\Phi_{CZ}(x,Q^2))}{R(\Phi_{asy}(x))}$ &
$\frac{R(\Phi_{BF}(x,Q^2))}{R(\Phi_{asy}(x))}$&
$\frac{R(\Phi_{BMS}(x,Q^2))}{R(\Phi_{asy}(x))}$ \\
\hline
  10 & 1.072  & 0.342 & 0.625 & 2.234\\ \hline
  35 & 4.238  & 1.115 & 4.074 & 0.976 \\ \hline
  95& 3.011  & 0.488 & 1,077 &0.561 \\ \hline
\end{tabular}
\end{center}
\caption{The distinction between $R(\Phi_{asy}(x))$ with
$R(\Phi_{i}(x,Q^{2}))$  (i=CLEO, CZ, BF, BMS) at c.m. energy $\sqrt
s=200\,\, GeV$.}\label{table3}
\end{table}

\newpage

\begin{table}[ht]
\begin{center}
\begin{tabular}{|c|c|c|c|c|c}\hline
$y$ & $\frac{r(\Phi_{CLEO}(x,Q^2))}{r(\Phi_{asy}(x))}$ &
$\frac{r(\Phi_{CZ}(x,Q^2))}{r(\Phi_{asy}(x))}$ &
$\frac{r(\Phi_{BF}(x,Q^2))}{r(\Phi_{asy}(x))}$ &
$\frac{R(\Phi_{BMS}(x,Q^2))}{R(\Phi_{asy}(x))}$ \\
\hline
  -2.52 & 5.002  & 0.823 & 3.148&6.294 \\ \hline
  -1.92 & 1.538  & 0.285 & 0.447 &1.089 \\ \hline
  0.78& 0.504  & 0.861 & 4.351 &4.149\\ \hline
\end{tabular}
\end{center}
\caption{The distinction between $R(\Phi_{asy}(x))$ with
$R(\Phi_{i}(x,Q^{2}))$ (i=CLEO, CZ, BF, BMS) at c.m. energy $\sqrt
s=200\,\, GeV$ and $p_T=15.5\,\, GeV/c$.} \label{table4}
\end{table}

\newpage

\begin{figure}[htb]
\vskip 1.2cm \epsfxsize 16cm \centerline{\epsfbox{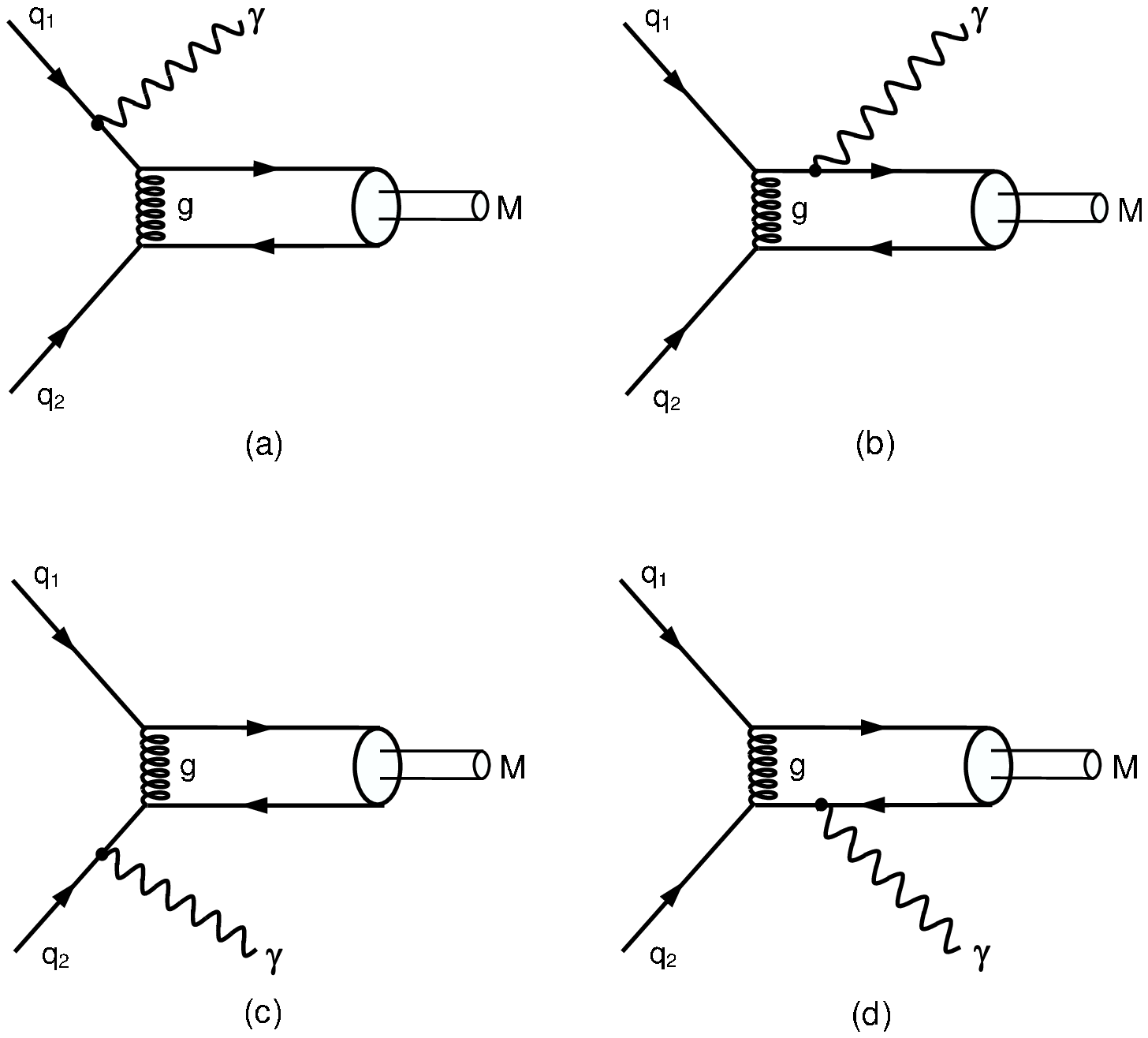}}
\vskip-5cm \caption{Feynman diagrams for the  higher-twist
subprocess, $q_1 q_2 \to \pi^{+}(or\,\,\pi^{-})\gamma.$}
\label{Fig1}
\end{figure}

\newpage

\begin{figure}[htb]
\vskip-1.2cm\epsfxsize 11.8cm \centerline{\epsfbox{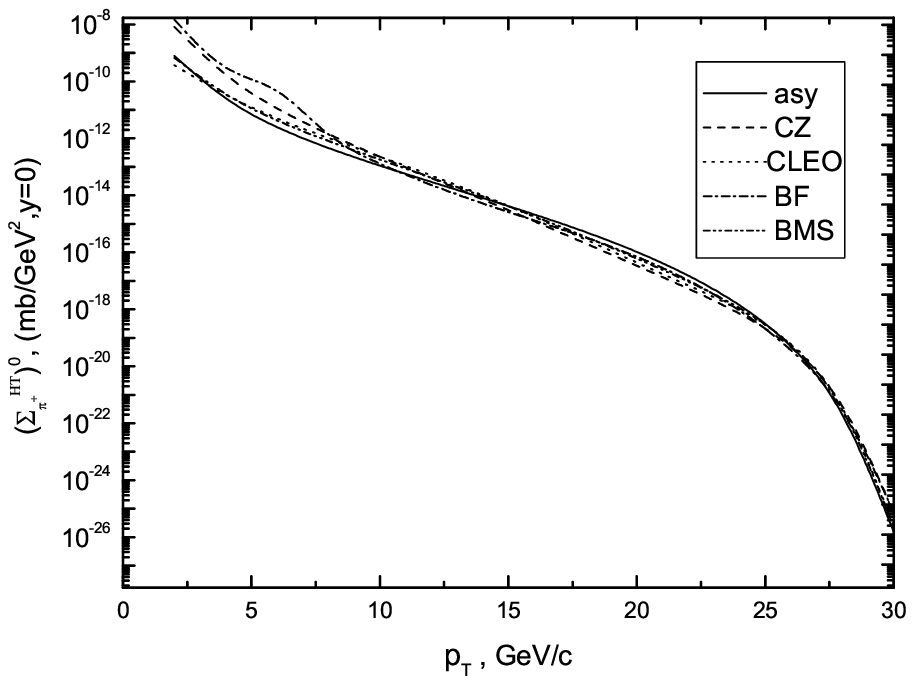}}
\vskip-0.2cm \caption{Higher-twist $\pi^{+}$ production cross
section $(\Sigma^{HT})^{0}$ as a function of the $p_{T}$ transverse
momentum of the pion at the c.m.energy  $\sqrt s=62.4\,\,
GeV$.}\label{Fig2}
 \vskip-1.0cm
\vskip 1.8cm \epsfxsize 11.8cm \centerline{\epsfbox{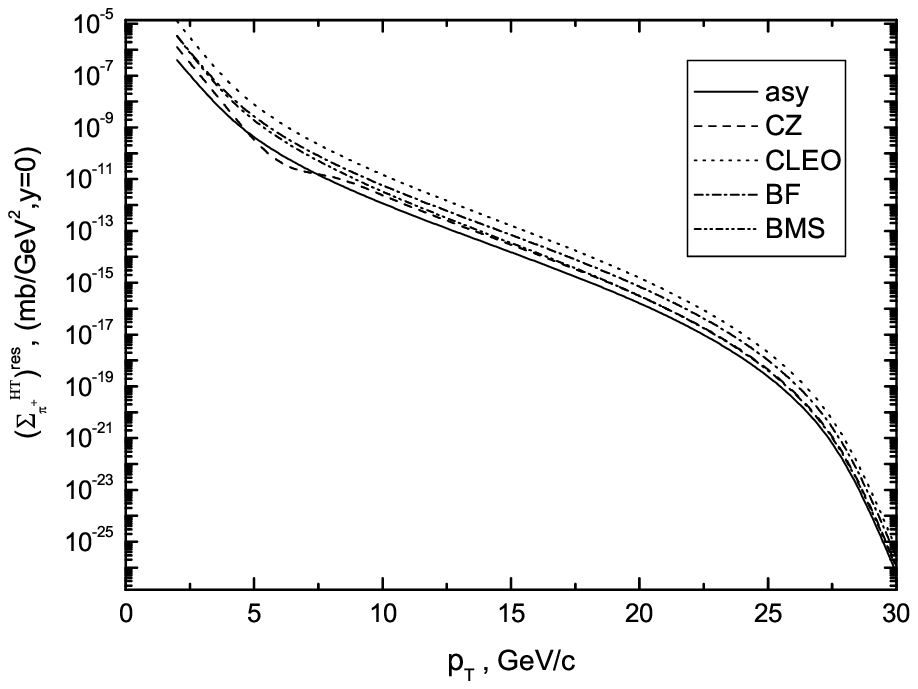}}
\vskip-0.05cm \caption{Higher-twist $\pi^{+}$ production cross
section $(\Sigma^{HT})^{res}$ as a function of the $p_{T}$
transverse momentum of the pion at the c.m.energy  $\sqrt s=62.4\,\,
GeV$.} \label{Fig3}
\end{figure}

\newpage

\begin{figure}[htb]
 \vskip-1.2cm\epsfxsize 11.8cm \centerline{\epsfbox{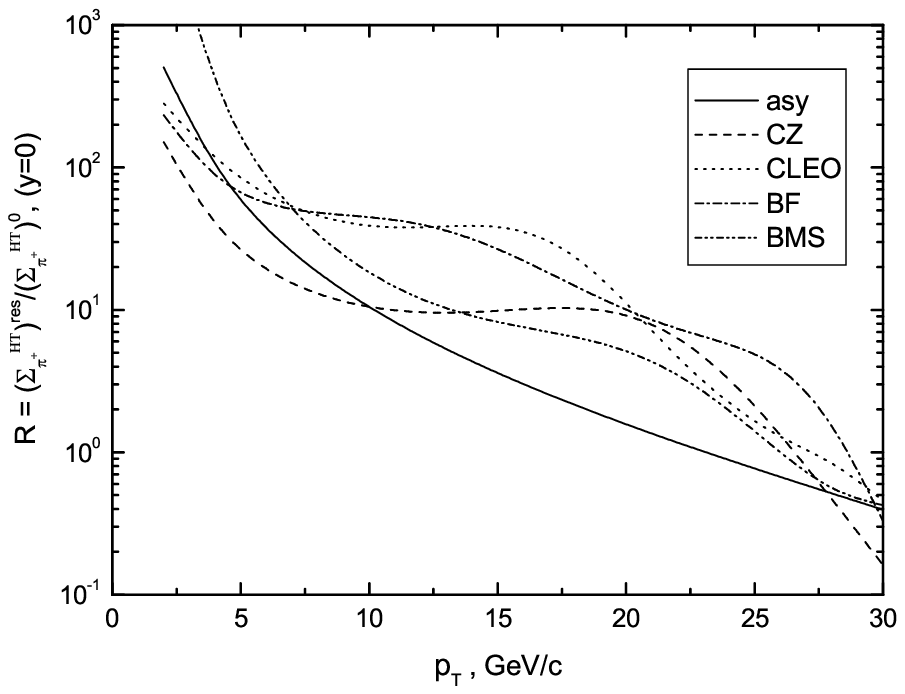}} \vskip-0.2cm
\caption{Ratio
$R=(\Sigma_{\pi^{+}}^{HT})^{res}/(\Sigma_{\pi^{+}}^{HT})^{0}$, where
higher-twist contribution are calculated for the pion rapidity $y=0$
at the c.m.energy $\sqrt s=62.4\,\,GeV$ as a function of the pion
transverse momentum, $p_{T}$.} \label{Fig4}
 \vskip 1.8cm
 \vskip-1.2cm\epsfxsize 11.8cm \centerline{\epsfbox{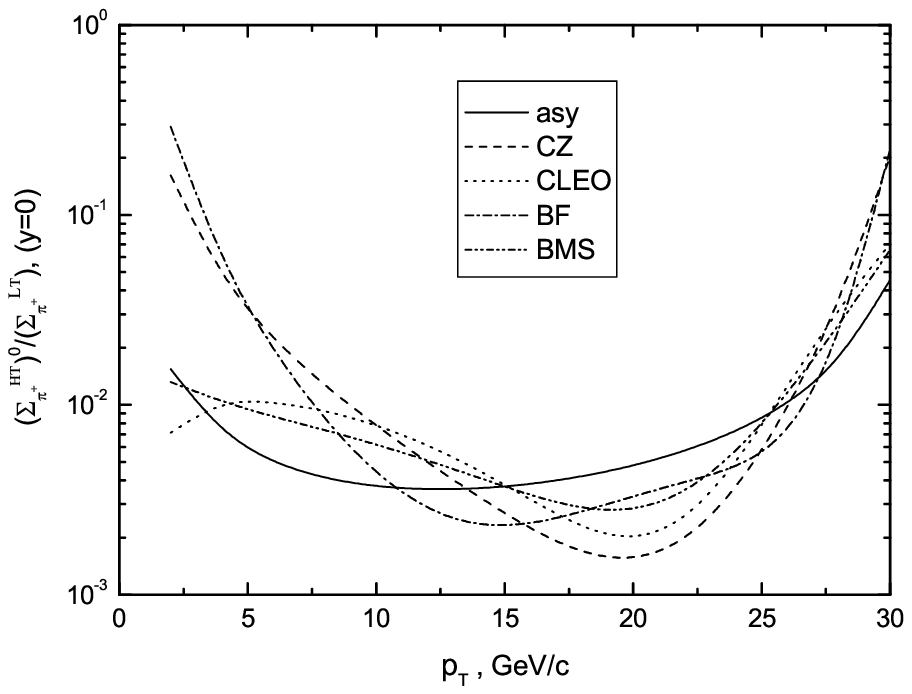}} \vskip-0.2cm
\caption{Ratio
$(\Sigma_{\pi^{+}}^{HT})^{0}/(\Sigma_{\pi^{+}}^{LT})$, as a function
of the  $p_{T}$ transverse momentum of the pion  at the c.m. energy
$\sqrt s=62.4\,\,GeV$.} \label{Fig5}
\end{figure}

\newpage

\begin{figure}[htb]
\vskip-1.2cm \epsfxsize 11.8cm \centerline{\epsfbox{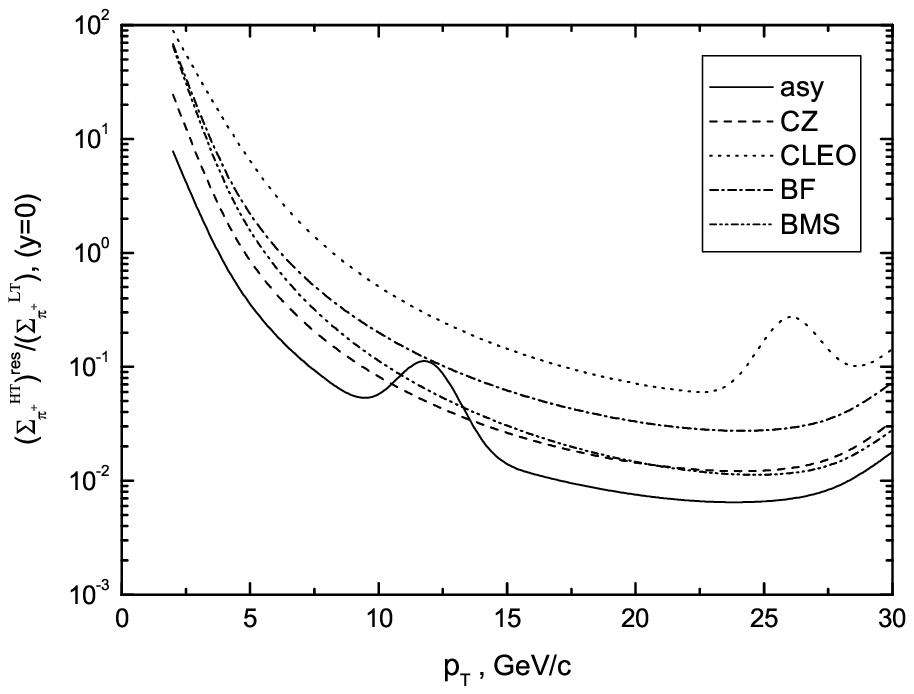}}
\vskip-0.2cm \caption{Ratio
$(\Sigma_{\pi^{+}}^{HT})^{res}/(\Sigma_{\pi^{+}}^{LT})$, as a
function  of the  $p_{T}$ transverse momentum of the pion  at the
c.m. energy $\sqrt {s} =62.4\,\,GeV$.} \label{Fig6}
\vskip-1.0cm
\vskip 1.8cm\epsfxsize 11.8cm \centerline{\epsfbox{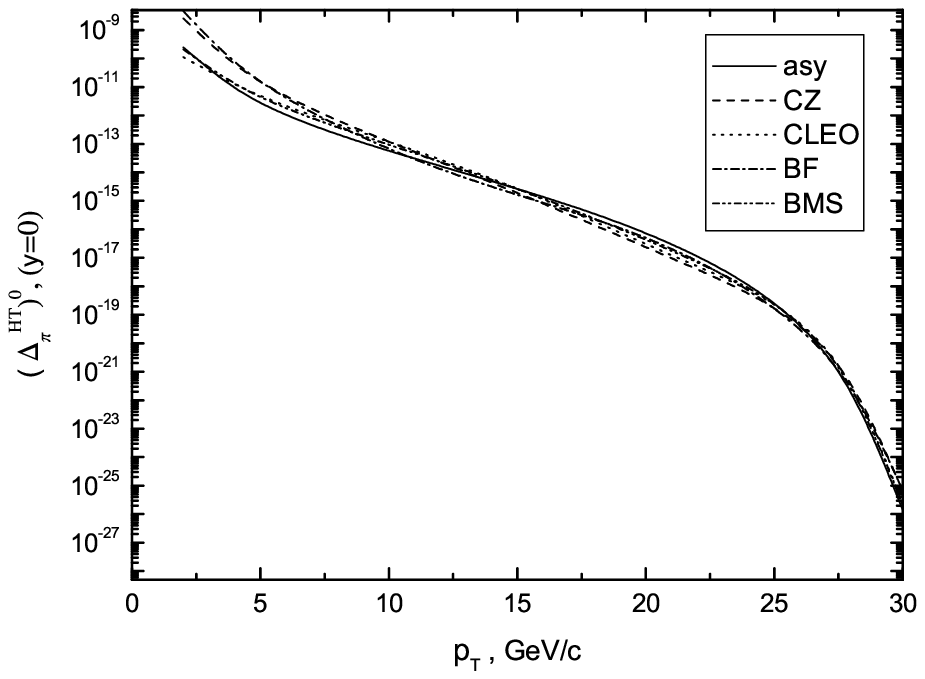}}
\vskip-0.05cm \caption{The difference of the higher-twist cross
section,
$(\Delta_{\pi}^{HT})^{0}=(\Sigma_{\pi^{+}}^{HT})^{0}-(\Sigma_{\pi^{-}}^{HT})^{0}$,
as a function of the pion transverse momentum, $p_{T}$, at the
c.m.energy $\sqrt s=62.4\,\, GeV$.}\label{Fig7}
\end{figure}

\newpage

\begin{figure}[htb]
\vskip-1.2cm \epsfxsize 11.8cm \centerline{\epsfbox{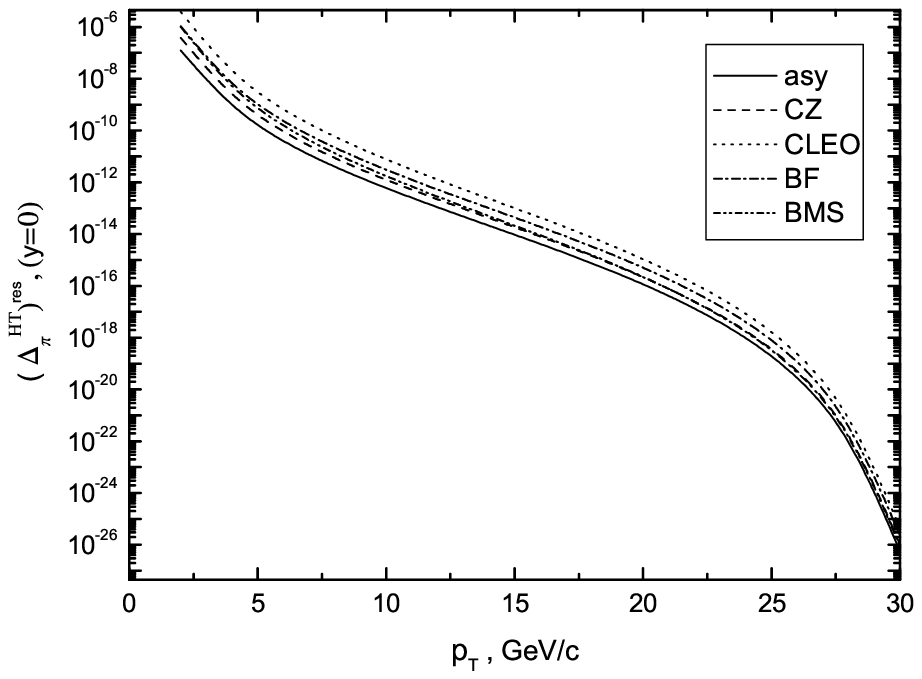}}
\vskip-0.2cm \caption{The difference of the higher-twist cross
section,
$(\Delta_{\pi}^{HT})^{res}=(\Sigma_{\pi^{+}}^{HT})^{res}-(\Sigma_{\pi^{-}}^{HT})^{res}$,
as a function of the pion transverse momentum, $p_{T}$, at the
c.m.energy $\sqrt s=62.4\,\, GeV$.} \label{Fig8}
\vskip-0.4cm
\vskip 0.8cm \epsfxsize 11.8cm \centerline{\epsfbox{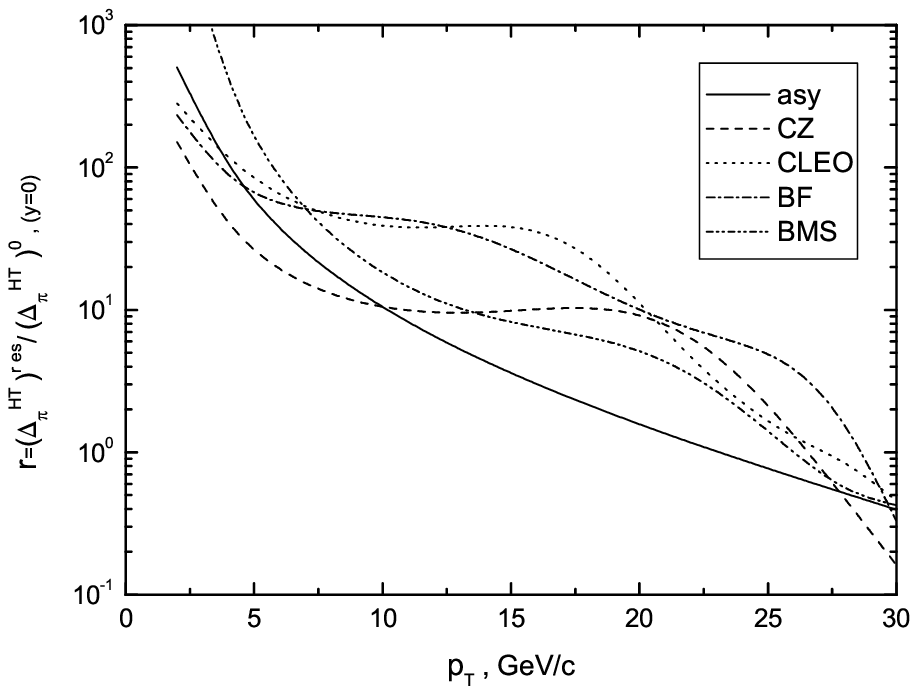}}
\vskip-0.2cm \caption{Ratio
$r=(\Delta_{\pi}^{HT})^{res}/(\Delta_{\pi}^{HT})^{0}$, where higher-
twist contributions are calculated for the pion rapidity $y=0$ at
the c.m. energy $\sqrt s=62.4\,\, GeV$, as a function of the pion
transverse momentum, $p_T$.} \label{Fig9}
\end{figure}

\newpage

\begin{figure}[htb]
\vskip-1.2cm \epsfxsize 11.8cm \centerline{\epsfbox{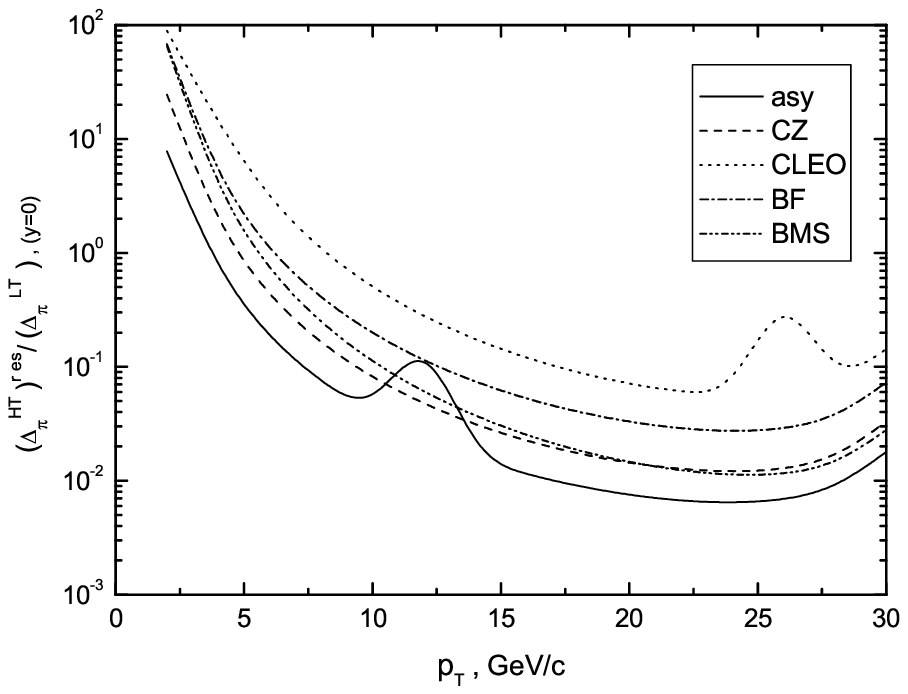}}
\vskip-0.2cm \caption{Ratio
$(\Delta_{\pi}^{HT})^{res}/(\Delta_{\pi}^{LT})$, where higher-twist
contributions are calculated for the pion rapidity $y=0$ at the c.m.
energy $\sqrt s=62.4\,\, GeV$, as a function of the pion transverse
momentum, $p_T$.} \label{Fig10} \vskip 1.8cm \epsfxsize 11.8cm
\centerline{\epsfbox{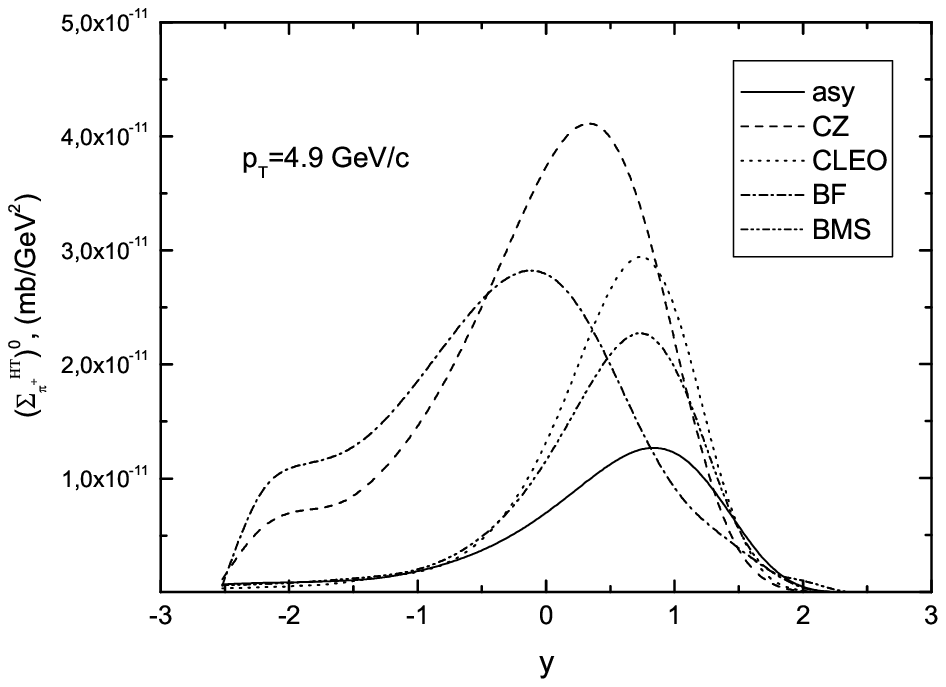}} \vskip-0.05cm \caption{Higher-
twist $\pi^{+}$ production cross section
$(\Sigma_{\pi^{+}}^{HT})^{0}$, as a function of the $y$ rapidity of
the pion at the  transverse momentum of the pion $p_T=4.9\,\,
GeV/c$, at the c.m. energy $\sqrt s=62.4\,\, GeV$.} \label{Fig11}
\end{figure}

\newpage

\begin{figure}[htb]
\vskip-1.2cm\epsfxsize 11.8cm \centerline{\epsfbox{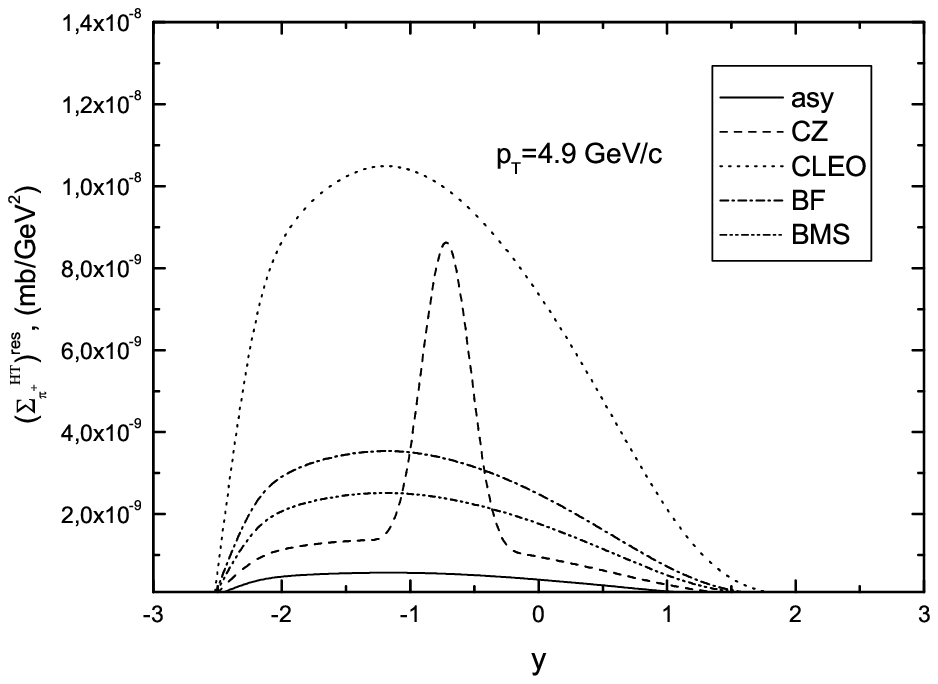}}
\vskip-0.2cm \caption{Higher-twist $\pi^{+}$ production cross
section $(\Sigma_{\pi^{+}}^{HT})^{res}$, as a function of the $y$
rapidity of the pion at the  transverse momentum of the pion
$p_T=4.9\,\, GeV/c$, at the c.m. energy $\sqrt s=62.4\,\, GeV$.}
\label{Fig12} \vskip 1.8cm
\epsfxsize 11.8cm \centerline{\epsfbox{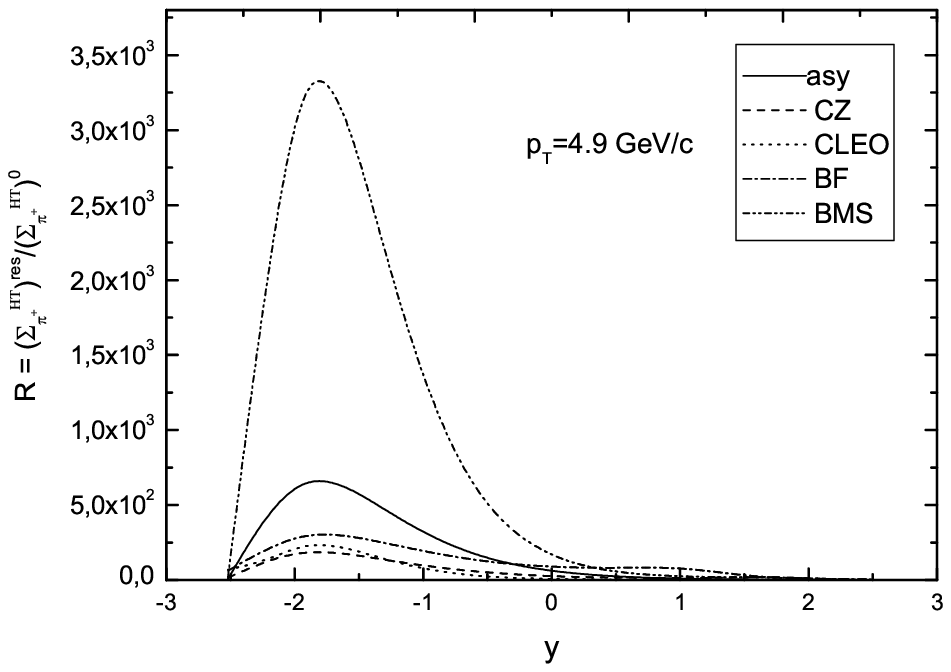}} \vskip-0.05cm
\caption{Ratio
$R=(\Sigma_{\pi^{+}}^{HT})^{res}/(\Sigma_{\pi}^{HT})^{0}$, as a
function of the $y$ rapidity of the pion at the  transverse momentum
of the pion $p_T=4.9\,\, GeV/c$, at the c.m. energy $\sqrt
s=62.4\,\, GeV$.} \label{Fig13}
\end{figure}

\newpage

\begin{figure}[htb]
\vskip-1.2cm\epsfxsize 11.8cm \centerline{\epsfbox{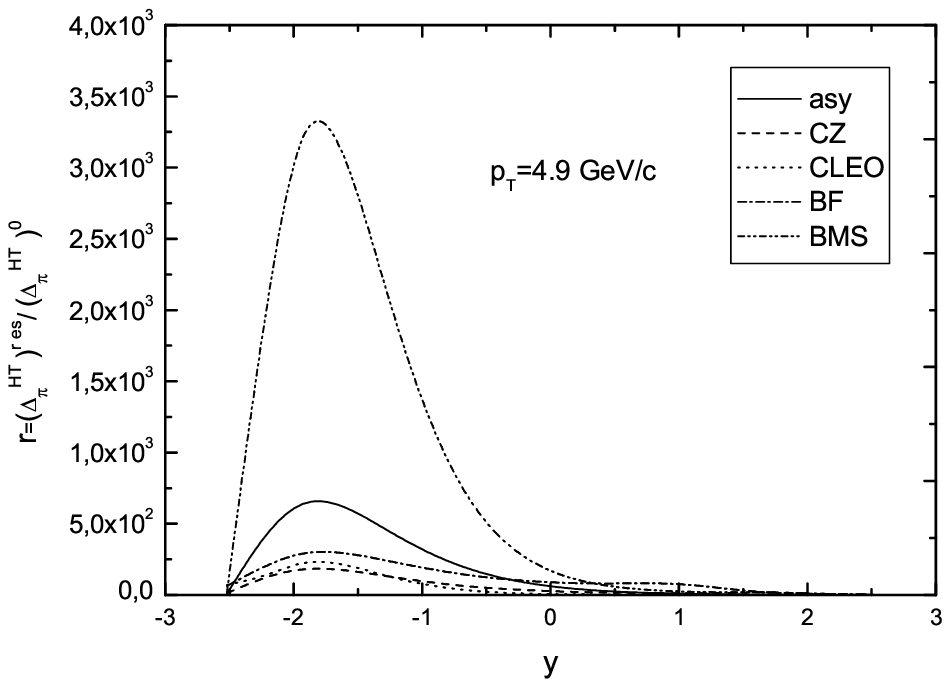}}
\vskip-0.2cm \caption{Ratio
$r=(\Delta_{\pi}^{HT})^{res}/(\Delta_{\pi}^{HT})^{0}$, as a function
of the $y$ rapidity of the pion at the  transverse momentum of the
pion $p_T=4.9\,\, GeV/c$, at the c.m. energy $\sqrt s=62.4\,\,
GeV$.} \label{Fig14} \vskip 1.8cm
\epsfxsize 11.8cm \centerline{\epsfbox{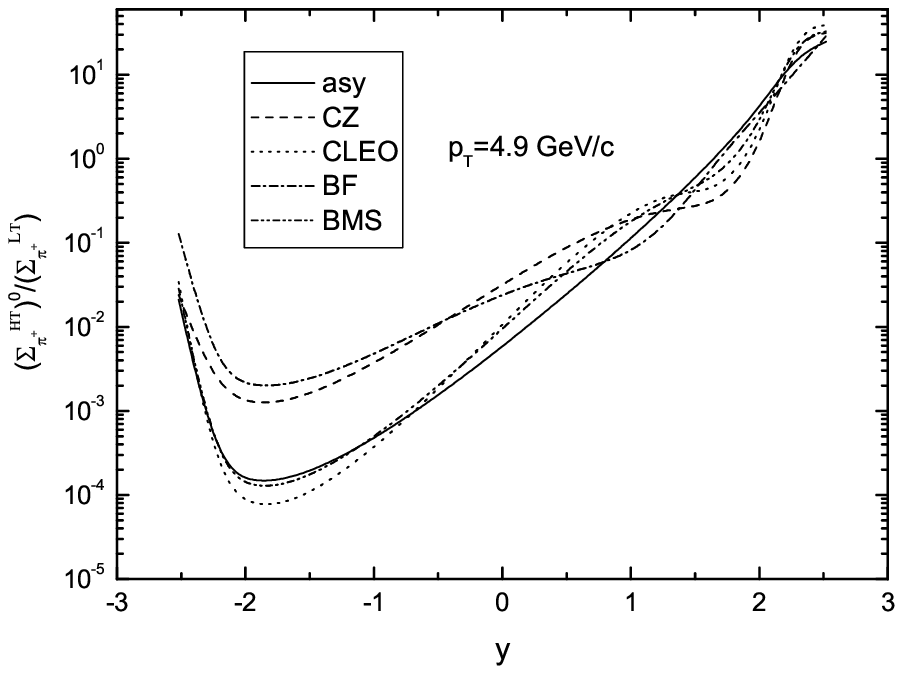}} \vskip-0.05cm
\caption{Ratio
$(\Sigma_{\pi^{+}}^{HT})^{0}/(\Sigma_{\pi^{+}}^{LT})$, as a function
of the $y$ rapidity of the pion at the  transverse momentum of the
pion $p_T=4.9\,\, GeV/c$, at the c.m. energy $\sqrt s=62.4\,\,
GeV$.} \label{Fig15}
\end{figure}

\newpage

\begin{figure}[htb]
\vskip-1.2cm \epsfxsize 11.8cm \centerline{\epsfbox{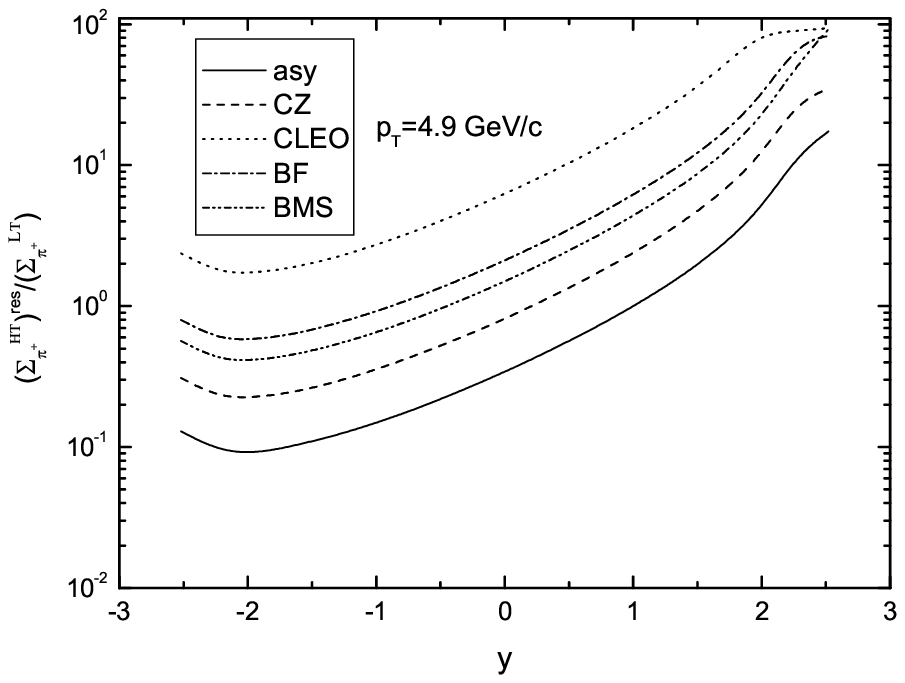}}
\vskip-0.2cm \caption{Ratio
$(\Sigma_{\pi^{+}}^{HT})^{res}/(\Sigma_{\pi^{+}}^{LT})$, as a
function of the $y$ rapidity of the pion at the  transverse
momentum of the pion $p_T=4.9\,\, GeV/c$, at the c.m. energy
$\sqrt s=62.4\,\, GeV$.} \label{Fig16} \vskip 1.8cm
\epsfxsize 11.8cm \centerline{\epsfbox{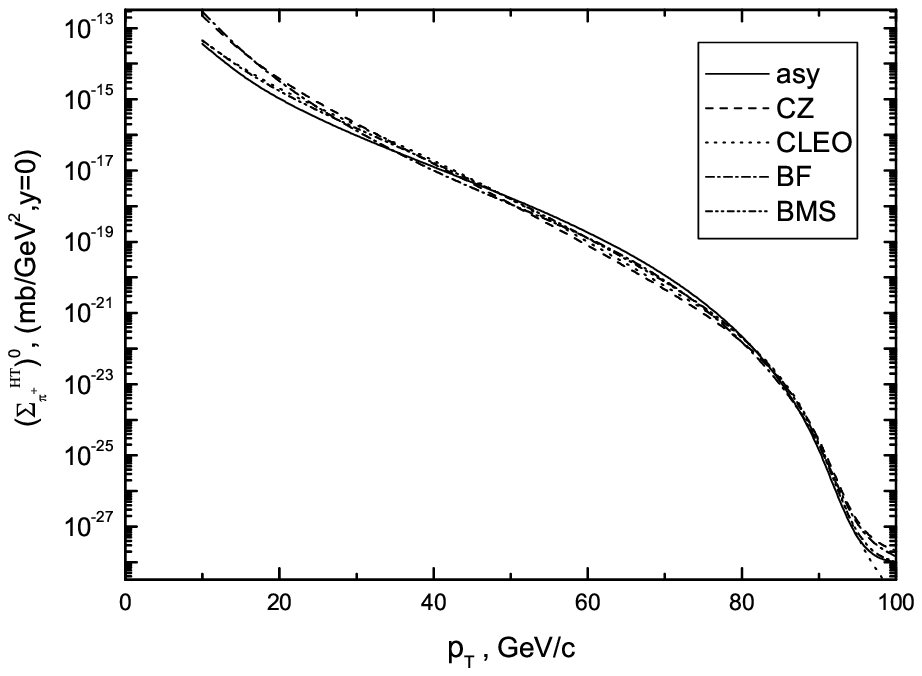}} \vskip-0.05cm
\caption{Higher-twist $\pi^{+}$ production cross section
$(\Sigma_{\pi^{+}}^{HT})^{0}$ as a function of the $p_{T}$
transverse momentum of the pion at the c.m.energy  $\sqrt s=200\,\,
GeV$.} \label{Fig17}
\end{figure}

\newpage

\begin{figure}[htb]
\vskip-1.2cm\epsfxsize 11.8cm \centerline{\epsfbox{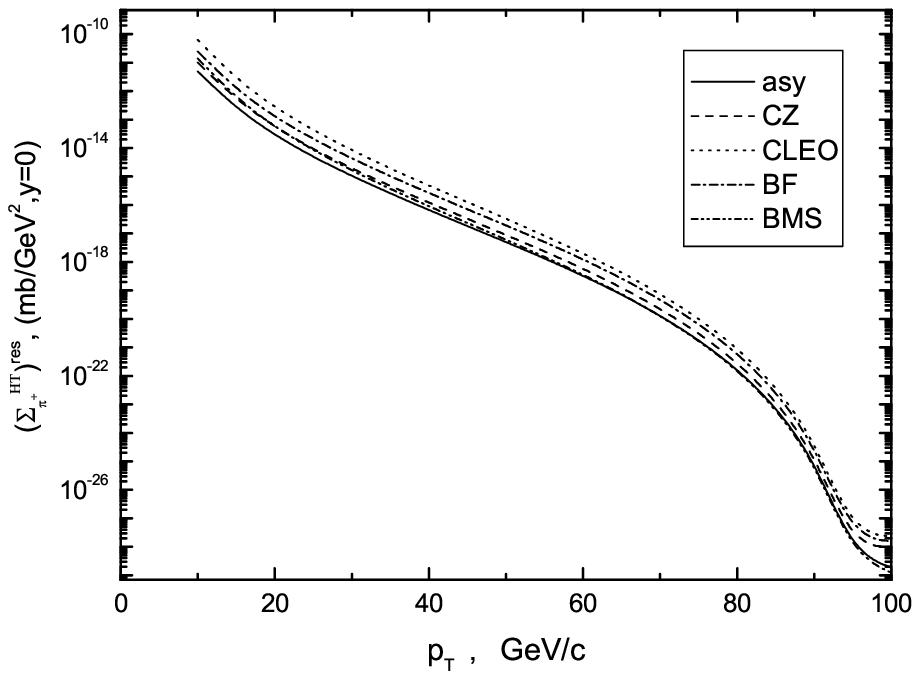}}
\vskip-0.2cm \caption{Higher-twist $\pi^{+}$ production cross
section $(\Sigma_{\pi^{+}}^{HT})^{res}$ as a function of the $p_{T}$
transverse momentum of the pion at the c.m.energy $\sqrt s=200\,\,
GeV$.} \label{Fig18} \vskip 1.8cm
\epsfxsize 11.8cm \centerline{\epsfbox{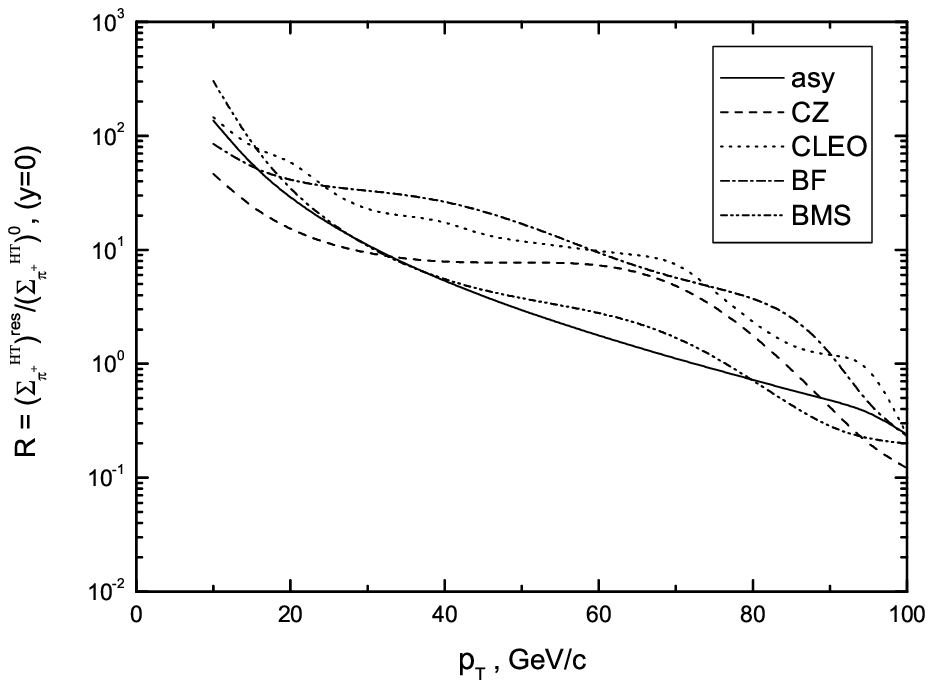}} \vskip-0.05cm
\caption{Ratio
$R=(\Sigma_{\pi^{+}}^{HT})^{res}/(\Sigma_{\pi^{+}}^{HT})^{0}$, where
higher-twist contribution are calculated for the pion rapidity $y=0$
at the c.m.energy $\sqrt s=200\,\,GeV$ as a function of the pion
transverse momentum, $p_{T}$.} \label{Fig19}
\end{figure}

\newpage

\begin{figure}[htb]
\vskip-1.2cm\epsfxsize 11.8cm \centerline{\epsfbox{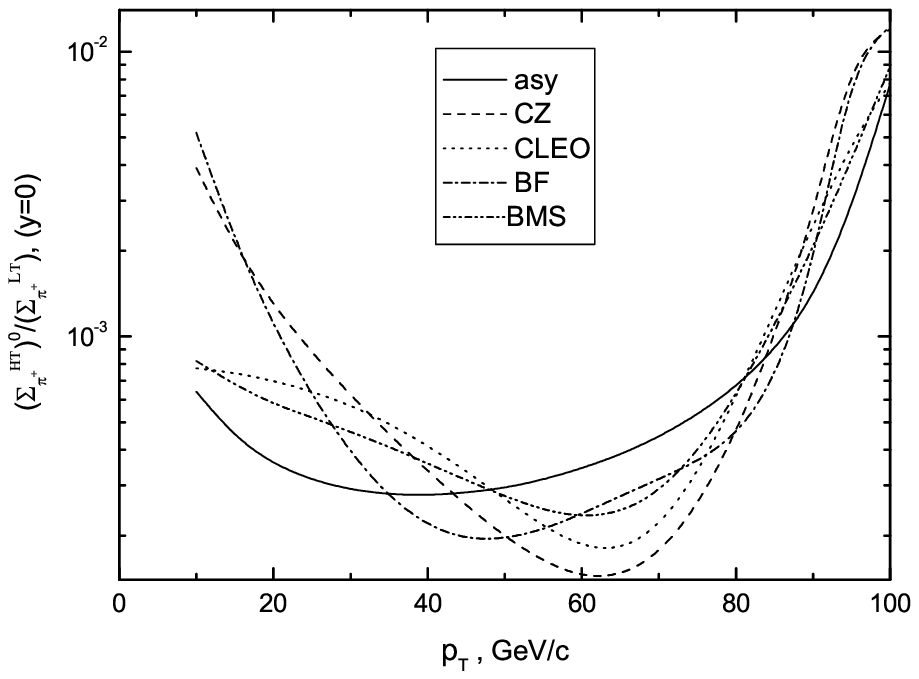}}
\vskip-0.2cm \caption{Ratio
$(\Sigma_{\pi^{+}}^{HT})^{0}/(\Sigma_{\pi^{+}}^{LT})$, as a function
of the  $p_{T}$ transverse momentum of the pion  at the c.m. energy
$\sqrt s=200\,\,GeV$.} \label{Fig20} \vskip 1.8cm
\epsfxsize 11.8cm \centerline{\epsfbox{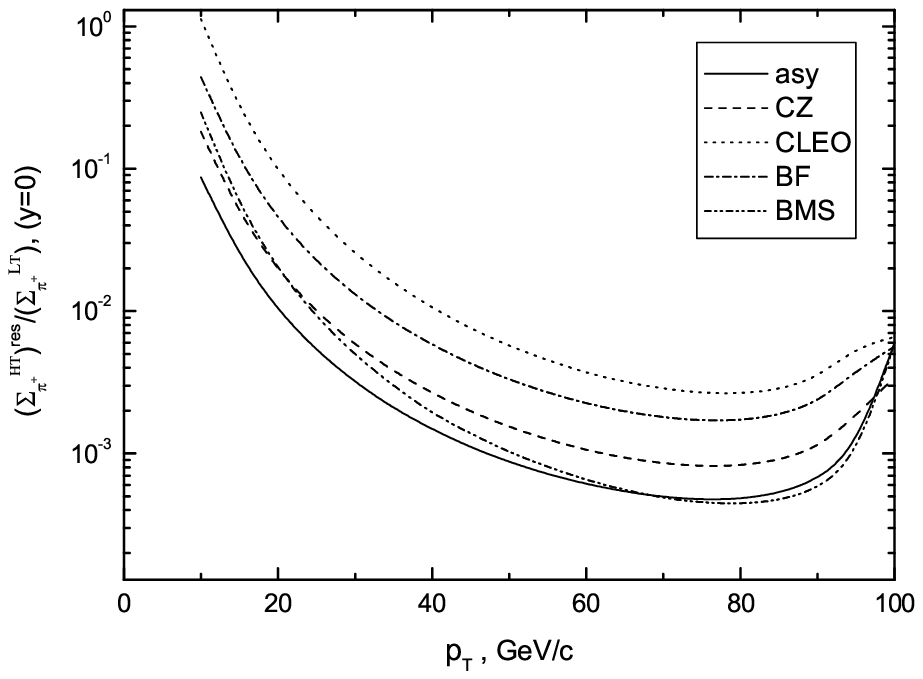}} \vskip-0.05cm
\caption{Ratio
$(\Sigma_{\pi^{+}}^{HT})^{res}/(\Sigma_{\pi^{+}}^{LT})$, as a
function  of the  $p_{T}$ transverse momentum of the pion  at the
c.m. energy $\sqrt s=200\,\,GeV$.} \label{Fig21}
\end{figure}

\newpage

\begin{figure}[htb]
\vskip-1.2cm\epsfxsize 11.8cm \centerline{\epsfbox{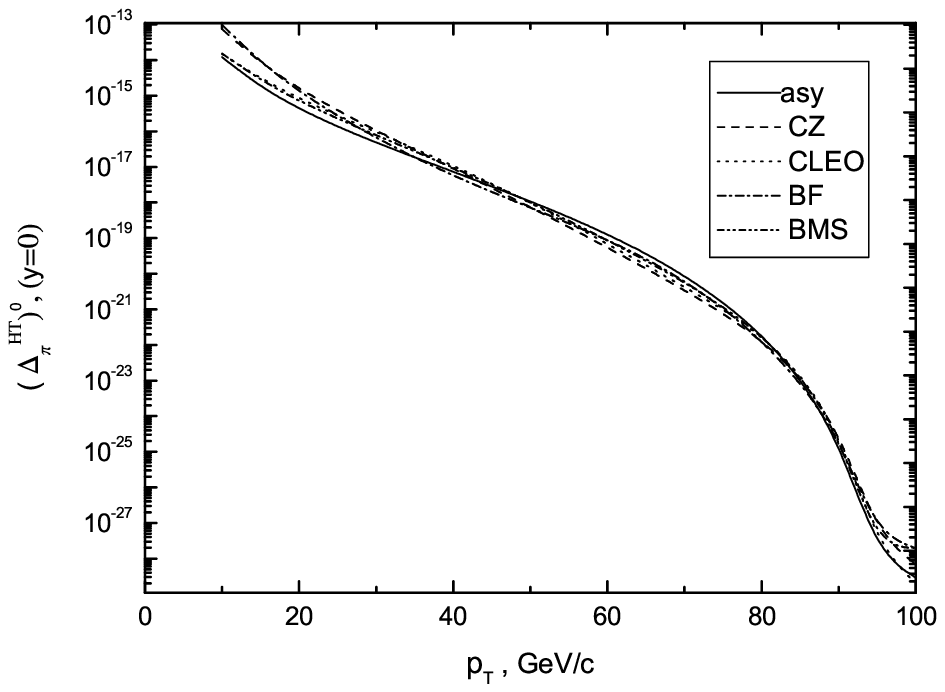}}
\vskip-0.2cm \caption{The difference of the higher-twist cross
section,
$(\Delta_{\pi}^{HT})^{0}=(\Sigma_{\pi^{+}}^{HT})^{0}-(\Sigma_{\pi^{-}}^{HT})^{0}$,
as a function of the pion transverse momentum, $p_{T}$, at the
c.m.energy $\sqrt s=200\,\, GeV$.} \label{Fig22} \vskip 1.8cm
\epsfxsize 11.8cm \centerline{\epsfbox{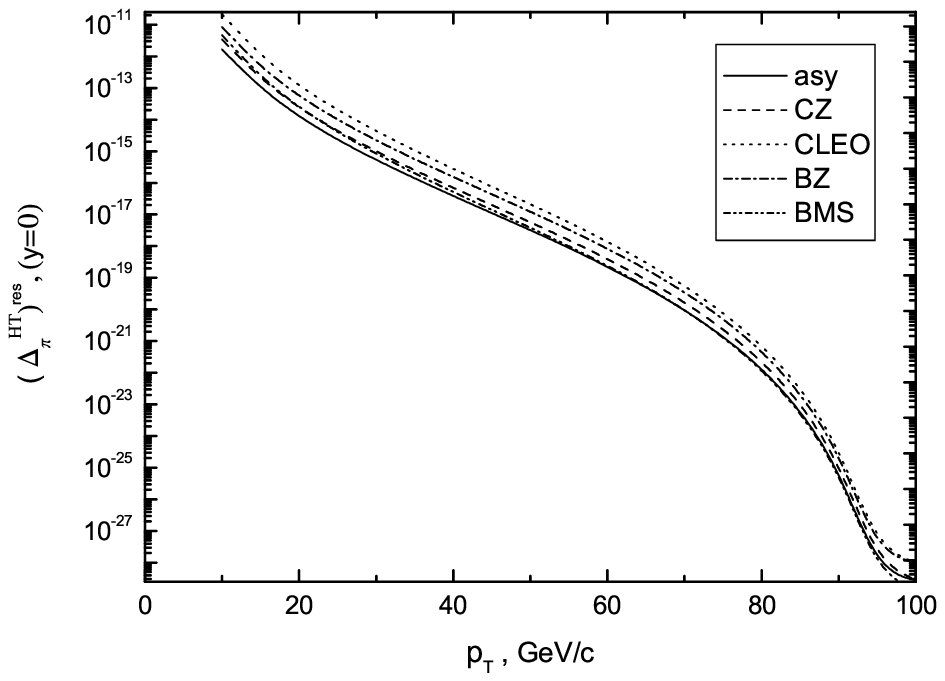}} \vskip-0.05cm
\caption{The difference of the higher-twist cross section,
$(\Delta_{\pi}^{HT})^{res}=(\Sigma_{\pi^{+}}^{HT})^{res}-(\Sigma_{\pi^{-}}^{HT})^{res}$,
as a function of the pion transverse momentum, $p_{T}$, at the
c.m.energy $\sqrt s=200\,\, GeV$.} \label{Fig23}
\end{figure}

\newpage

\begin{figure}[htb]
\vskip-1.2cm\epsfxsize 11.8cm \centerline{\epsfbox{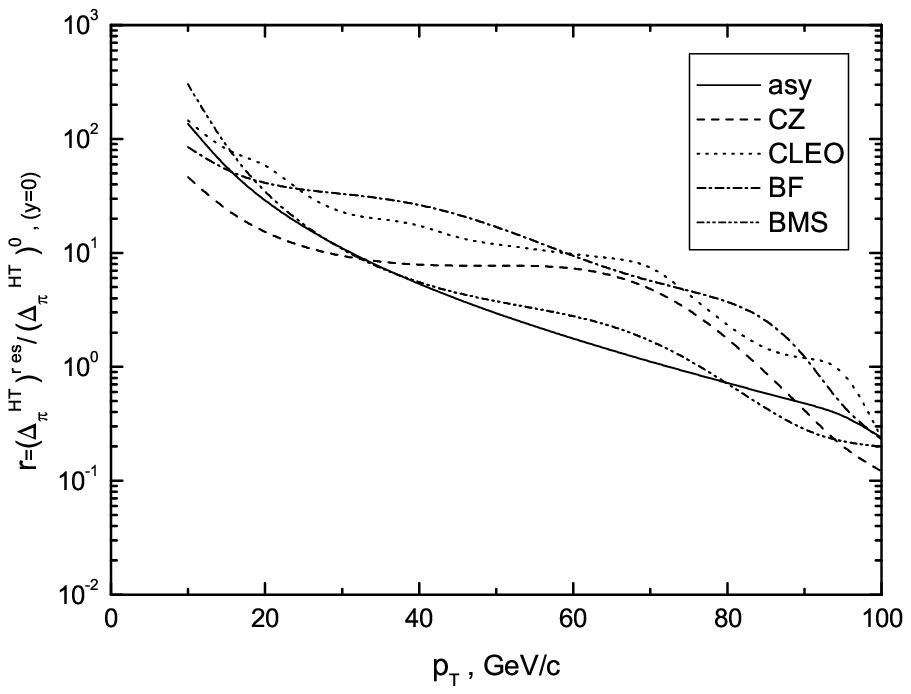}}
\vskip-0.2cm \caption{Ratio
$r=(\Delta_{\pi}^{HT})^{res}/(\Delta_{\pi}^{HT})^{0}$, where higher-
twist contributions are calculated for the pion rapidity $y=0$ at
the c.m. energy $\sqrt s=200\,\, GeV$, as a function of the pion
transverse momentum, $p_T$.} \label{Fig24} \vskip 1.8cm
\epsfxsize 11.8cm \centerline{\epsfbox{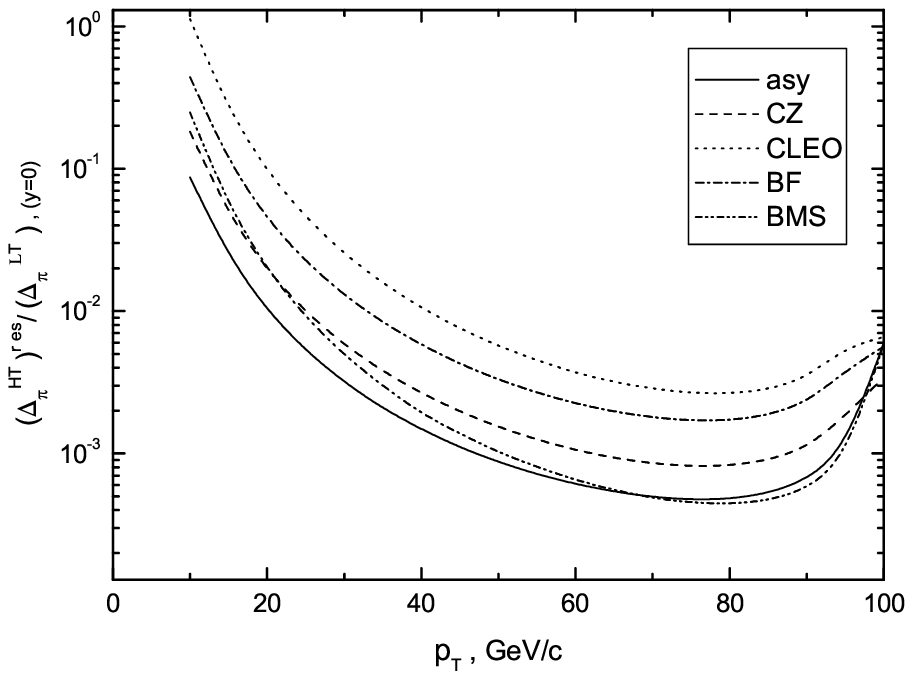}} \vskip-0.05cm
\caption{Ratio $(\Delta_{\pi}^{HT})^{res}/(\Delta_{\pi}^{LT})$,
where higher-twist contributions are calculated for the pion
rapidity $y=0$ at the c.m. energy $\sqrt s=200\,\, GeV$, as a
function of the pion transverse momentum, $p_T$.} \label{Fig25}
\end{figure}

\newpage

\begin{figure}[htb]
\vskip-1.2cm\epsfxsize 11.8cm \centerline{\epsfbox{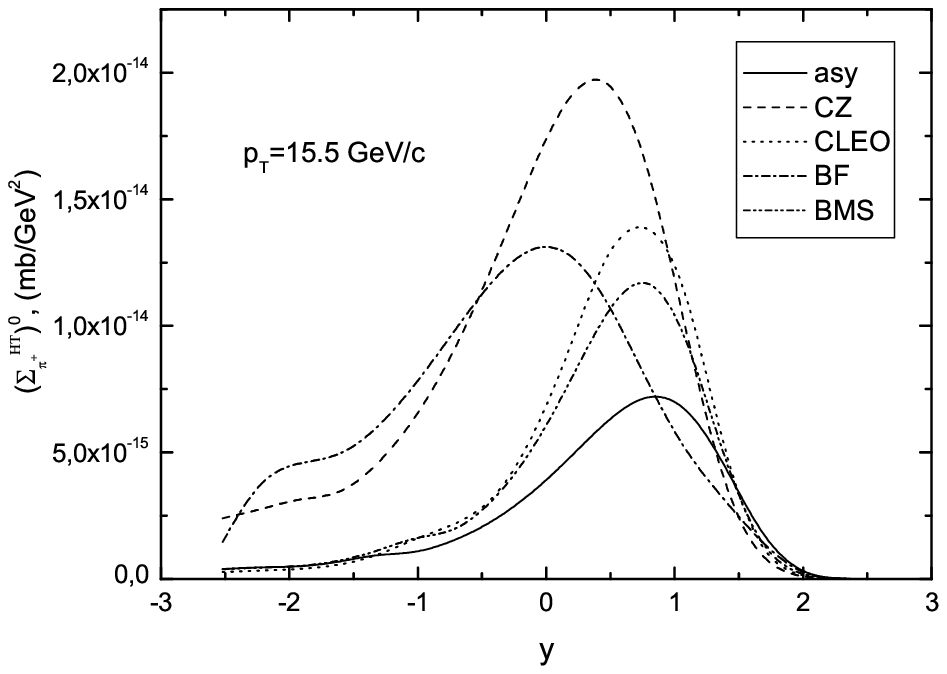}}
\vskip-0.2cm \caption{Higher-twist $\pi^{+}$ production cross
section $(\Sigma_{\pi^{+}}^{HT})^{0}$, as a function of the $y$
rapidity of the pion at the  transverse momentum of the pion
$p_T=15.5\,\, GeV/c$, at the c.m. energy $\sqrt s=200\,\, GeV$.}
\label{Fig26} \vskip 1.8cm
\epsfxsize 11.8cm \centerline{\epsfbox{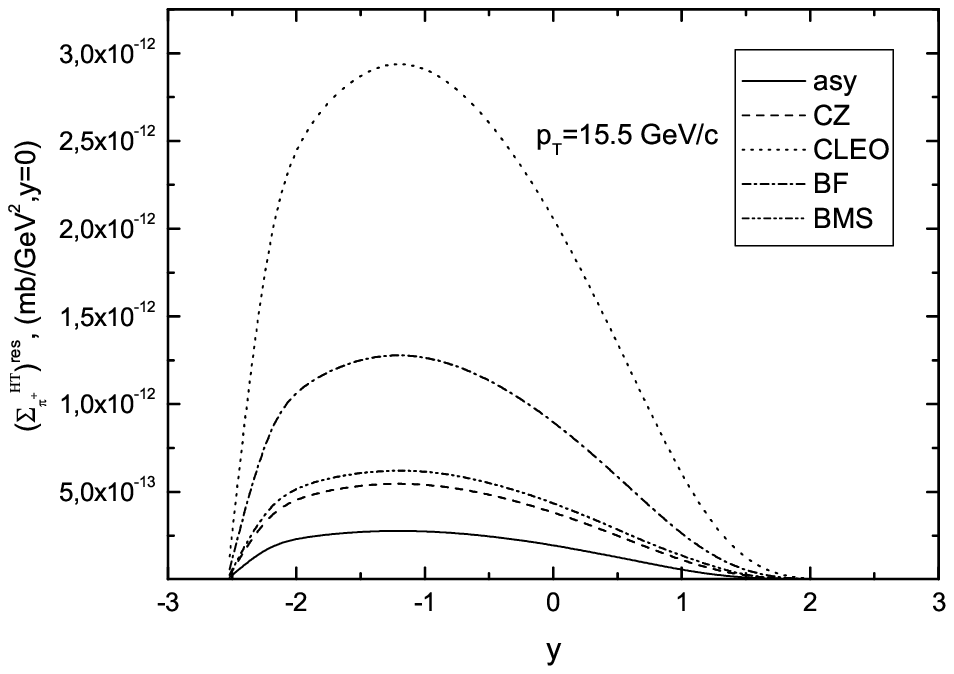}} \vskip-0.05cm
\caption{Higher-twist $\pi^{+}$ production cross section
$(\Sigma_{\pi^{+}}^{HT})^{res}$, as a function of the $y$ rapidity
of the pion at the  transverse momentum of the pion $p_T=15.5\,\,
GeV/c$, at the c.m. energy $\sqrt s=200\,\, GeV$.} \label{Fig27}
\end{figure}

\newpage

\begin{figure}[htb]
\vskip-1.2cm\epsfxsize 11.8cm \centerline{\epsfbox{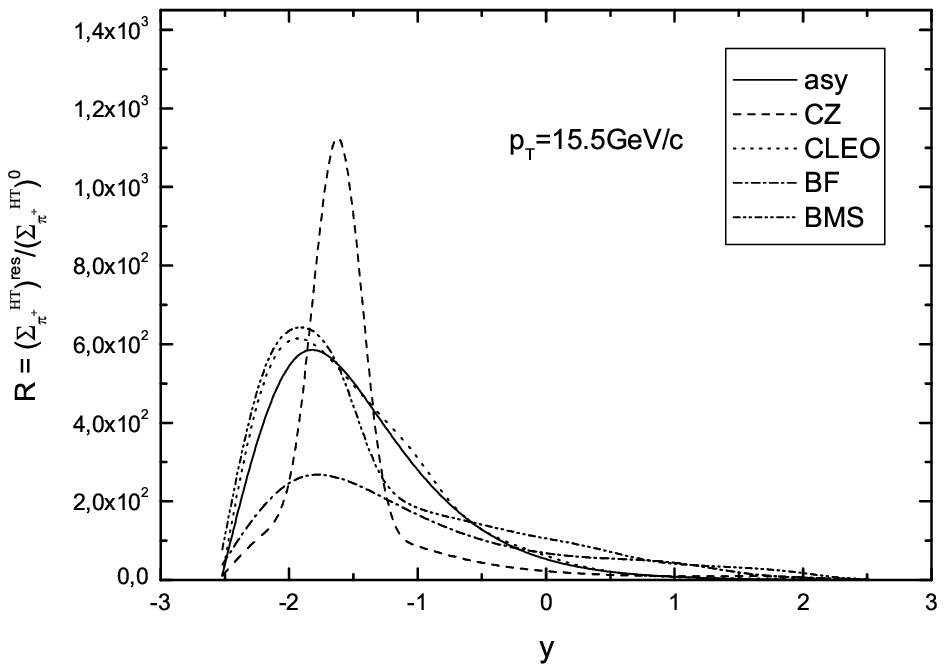}}
\vskip-0.2cm \caption{Ratio
$R=(\Sigma_{\pi^{+}}^{HT})^{res}/(\Sigma_{\pi}^{HT})^{0}$, as a
function of the $y$ rapidity of the pion at the  transverse momentum
of the pion $p_T=15.5\,\, GeV/c$, at the c.m. energy $\sqrt
s=200\,\, GeV$.} \label{Fig28} \vskip 1.8cm
\epsfxsize 11.8cm \centerline{\epsfbox{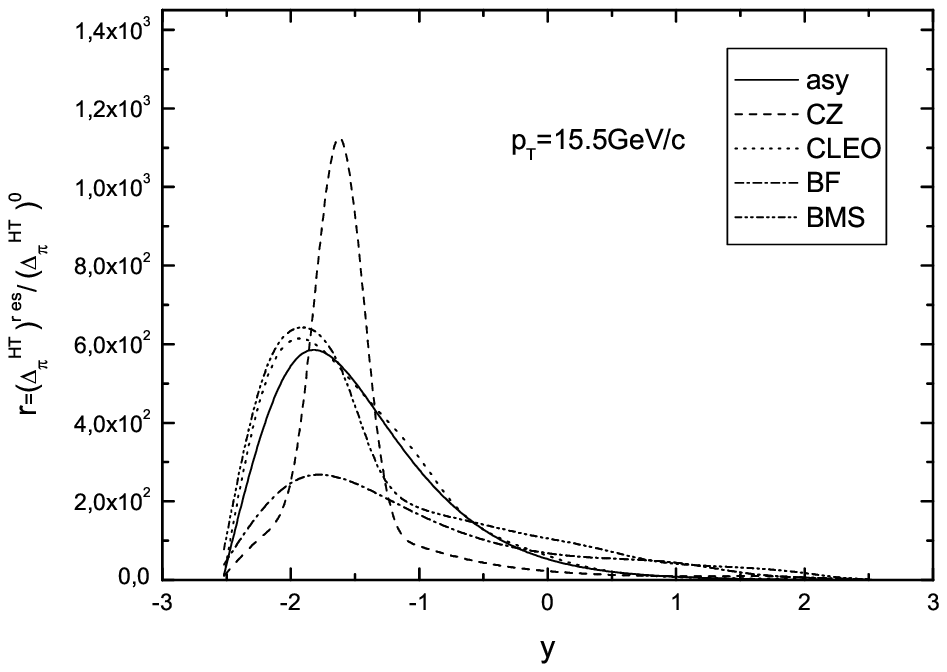}} \vskip-0.05cm
\caption{Ratio
$r=(\Delta_{\pi}^{HT})^{res}/(\Delta_{\pi}^{HT})^{0}$, as a
function of the $y$ rapidity of the pion at the  transverse
momentum of the pion $p_T=15.5\,\, GeV/c$, at the c.m. energy
$\sqrt s=200\,\, GeV$.} \label{Fig29}
\end{figure}

\newpage

\begin{figure}[htb]
\vskip-1.2cm\epsfxsize 11.8cm \centerline{\epsfbox{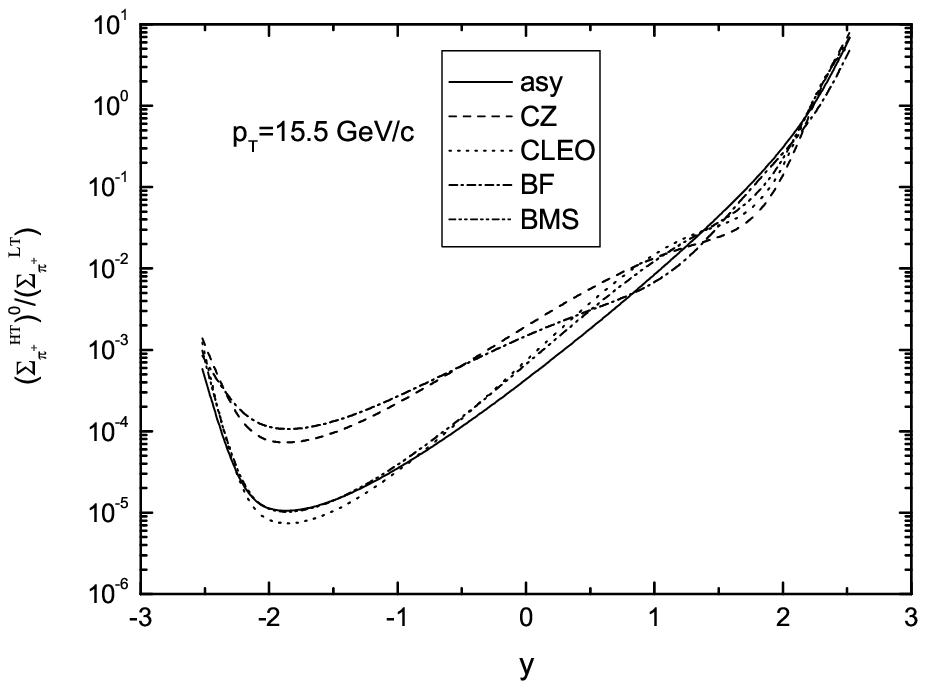}}
\vskip-0.2cm \caption{Ratio
$(\Sigma_{\pi^{+}}^{HT})^{0}/(\Sigma_{\pi^{+}}^{LT})$, as a function
of the $y$ rapidity of the pion at the  transverse momentum of the
pion $p_T=15.5\,\, GeV/c$, at the c.m. energy $\sqrt s=200\,\,
GeV$.} \label{Fig30} \vskip 1.8cm
\epsfxsize 11.8cm \centerline{\epsfbox{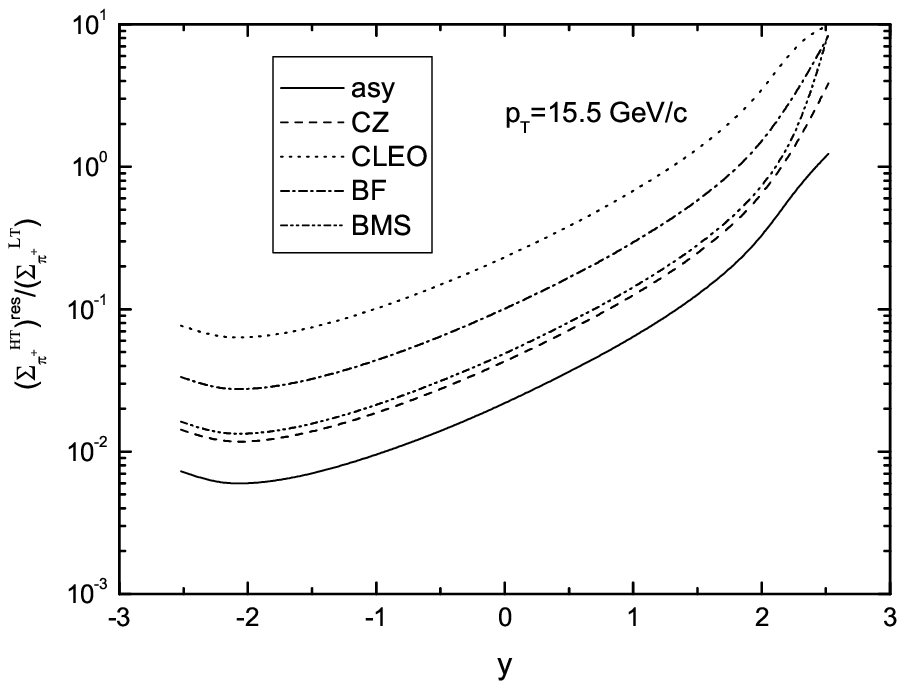}} \vskip-0.05cm
\caption{Ratio
$(\Sigma_{\pi^{+}}^{HT})^{res}/(\Sigma_{\pi^{+}}^{LT})$, as a
function of the $y$ rapidity of the pion at the  transverse momentum
of the pion $p_T=15.5\,\, GeV/c$, at the c.m. energy $\sqrt
s=200\,\, GeV$.} \label{Fig31}
\end{figure}

\end{document}